\documentclass[a4paper,11pt]{article}
\pdfoutput=1
\usepackage{jheppub}
\usepackage{amsmath,amssymb,amsfonts,amsthm,dsfont}

\title{Conformal Symmetry for Black Holes in Four Dimensions and Irrelevant Deformations}

 \author{Marco Baggio,}
 \author{Jan de Boer,}
 \author{Juan I. Jottar,}
 \author{and Daniel R. Mayerson}
 \affiliation{Institute for Theoretical Physics, University of Amsterdam,\\Science Park 904, Postbus 94485, 1090 GL Amsterdam, The Netherlands}

\abstract{It has been argued several times in the past that the structure of the entropy formula for general non-extremal asymptotically flat black holes in four dimensions can be understood in terms of an underlying conformal symmetry. A recent implementation of this idea, carried out by Cveti\v c and Larsen,
involves the replacement of a conformal factor in the original geometry by an alternative conformal factor in such a way that the near-horizon behavior and thermodynamic properties of the black hole remain unchanged, while only the asymptotics or ``environment" of the geometry are modified. The solution thus obtained, dubbed ``subtracted geometry", uplifts to an asymptotically AdS$_{3}\times S^{2}$ black hole in five dimensions, and an AdS/CFT interpretation is then possible. Building on this intuition we show that, at least in the static case, the replacement of the conformal factor can be implemented dynamically by means of an interpolating flow which we construct explicitly. Furthermore, we show that this flow can be understood as the effect of irrelevant perturbations from the point of view of the dual two-dimensional CFT, and we identify the quantum numbers of the operators responsible for the flow. This allows us to address quantitatively the validity of CFT computations for these asymptotically flat black holes and provides a framework
to systematically compute corrections to the CFT results.
}

\begin{document}
\maketitle
\flushbottom

\newpage
\section{Introduction}
Black holes and especially their entropy have always been a source of mystery in physics and, despite decades of effort and remarkable results on this topic, a full understanding of their properties is still lacking. The first microscopic understanding of black hole entropy was achieved for supersymmetric three-charge black holes, where the entropy formula was reproduced using an effective CFT description of the low-energy degrees of freedom \cite{Strominger:1996sh}. Later developments have expanded upon this description for other extremal and near-extremal black holes \cite{David:2002wn,Kraus:2006wn,Sen:2007qy}. The Kerr/CFT program \cite{Guica:2008mu,Hartman:2008pb,Compere:2012jk} has further provided indications for a CFT description of the low-energy physics for (near-)extremal rotating black holes.

Even when one identifies some sort of conformal symmetry in black hole related backgrounds, this does not need to have immediate physical implications,
nor does it have to imply the existence of a CFT whose physics is relevant for the black hole. The connection between the CFT and the black hole can
be purely numerological, purely kinematical, approximately dynamical, or exactly dynamical. An example of the latter is the BTZ black hole, whose
physics is completely captured by that of a $2d$ CFT. When there is an approximate dynamical equivalence the physical quantities of interest may or may not,
depending on the question, be approximately reproduced using CFT computations.\footnote{In the purely kinematical case, it is only the near horizon AdS${}_2$ geometry of an extremal black hole which agrees with the near horizon geometry of a BTZ black hole, and in the numerological case one only has a match
between the CFT entropy and the black hole entropy.} In this paper we will mostly be interested in this case.

Unfortunately, a systematic approach to find the approximate dynamical CFT description for the low-energy physics of black holes is still lacking. In general, the low energy near-horizon modes of general black holes (as opposed to (near-)extremal ones) do not seem to decouple from the asymptotics of the geometry, thus effectively preventing one from being able to take a decoupling limit and find an effective CFT description of these modes. Even so, there has been much interest in the fact that the massless wave equation in a (general) black hole background admits a $SL(2,\mathds{R})$ symmetry when certain terms are removed. The offending terms are indeed small and can be neglected in certain limits (near-extremal, near extreme rotating, low energy \cite{Cvetic:1997uw,Maldacena:1996ix,Guica:2008mu,Cvetic:2009jn,Castro:2010fd}). However reminiscent of conformal symmetry this approximate $SL(2,\mathds{R})$ symmetry may be, the terms ``breaking'' this symmetry are not small for general black holes and thus can not justifiably be ignored. The program of ``hidden conformal symmetry'' asserts that the conformal symmetry is there after all, but it is spontaneously broken \cite{Castro:2010fd}.

A recent development by Cveti\v c and Larsen \cite{Cvetic:2011hp,Cvetic:2011dn}, inspired by the approximate $SL(2,\mathds{R})$ symmetry of the massless scalar wave equation, provides further evidence for an approximate CFT description of general black holes in four and five dimensions far from extremality. They construct a so-called ``subtracted'' geometry, where the warp factor of the geometry is modified. Thus, the asymptotics of the black hole are changed from asymptotically flat to asymptotically conical \cite{Cvetic:2012tr}, but the thermodynamic properties of the black hole are left untouched. This subtracted geometry can then intuitively be thought of as ``putting the black hole in a box". This subtracted geometry, in addition to providing an exact $SL(2,\mathds{R})$ symmetry of the wave equation, can be uplifted one dimension higher to a geometry that is locally a product of AdS$_3$ and a two-sphere. Thus, a $2d$ CFT description of this black hole subtracted geometry is immediately obvious.

What is less obvious is the relation between the subtracted geometry and the original, asymptotically flat one, and in particular how this relation would be visible in the CFT description. Further developments \cite{Cvetic:2012tr} have made some progress in this direction by showing that the subtracted geometry can be obtained as a scaling limit of the original geometry. In this paper, we wish to address this problem and provide further evidence of the $2d$ CFT description of general asymptotically flat, non-rotating, four-charge black holes in four dimensions. If the $2d$ CFT description of the subtracted geometry is related to some IR limit of the asymptotic flat original geometry, then it is natural to expect this original geometry to be described by the CFT plus some irrelevant deformation.

We first construct a family of black hole solutions with various asymptotics in the STU model in four dimensions, of which both the subtracted geometry and the original geometry are members. With this explicit family of solutions, it is easy to find the linear perturbations of the subtracted geometry which must be turned on to flow to the original geometry. These perturbations can then be uplifted to perturbations of AdS$_3\times S^2$, where the standard AdS/CFT dictionary allows us to interpret and translate them as irrelevant deformations of the dual $2d$ CFT. We also explicitly determine the sources for these dimension $(2,2)$ operators in the CFT: in addition to providing further insight into the CFT description of asymptotically flat black holes, the explicit construction of these irrelevant operators and their sources allows us to quantitatively determine the limit of the validity of the CFT description.

The rest of the paper is organized as follows. In section \ref{sec:STU} we review the STU model and construct a four-parameter family of four-charged, non-rotating black holes with different asymptotics but the same thermodynamics, 
and explicitly identify the original (asymptotically flat) and subtracted geometry as members of this family. Then, in section \ref{sec:flows}, we perform a linear analysis to determine the perturbations needed to flow from the subtracted to the original geometry. In section \ref{sec:uplift} we uplift the subtracted geometry and the linear perturbations thereof to $5d$; we then consistently reduce to an effective three-dimensional description to easily identify the irrelevant operators and sources through the standard AdS/CFT dictionary. The determination of the sources gives us a clear criterion for the window in which the effective IR CFT description is valid. Finally, in section \ref{sec:discussion}, we summarize and discuss our findings. Various details of our calculations and useful formulae are collected in the appendix.

\section{The STU model}\label{sec:STU}
The STU model is a four-dimensional $\mathcal{N}=2$ supergravity theory coupled to three vector multiplets \cite{Cremmer:1984hj,Duff:1995sm,Behrndt:1996hu}. Its Lagrangian is given by\footnote{We mostly follow the notation and conventions of reference \cite{Virmani:2012kw}, which we found particularly useful.}
\begin{align}
\mathcal{L}_4 ={}&
  R {\, \star_4\mathds{1}} - \frac{1}{2}H_{ij} {\,\star_4 d h^i} \wedge
d h^j  - \frac{3}{2f^2}{\,\star_4 df} \wedge df \nonumber  - \frac{f^3}{2}{\,\star_4
 F^0} \wedge   F^0
\nonumber \\
&
 - \frac{1}{2 f^2}H_{ij}{\,\star_4 d \chi^i} \wedge d \chi^j  - \frac{f}{2}
H_{ij}{\, \star_4}\left( F^i +\chi^i  F^0\right) \wedge \left(F^j +\chi^j F^0\right)
\label{STUaction}
\\
&
+\frac{1}{2}C_{ijk}\,\chi^i   F^j  \wedge  F^k  +
\frac{1}{2} C_{ijk}\,\chi^i \chi^j  F^0  \wedge  F^k + \frac{1}{6}C_{ijk}\, \chi^i \chi^j \chi^k  F^0 \wedge   F^0\, ,
\nonumber
\end{align}
\noindent where the fields $f$ and $h^i$ ($i=1,2,3$) are scalars, $\chi^{i}$ are pseudoscalars, and $F^{0}$ and $F^{i}$ are $U(1)$ gauge field strengths. The metric $H_{ij}$ on the scalar moduli space is diagonal with entries $H_{ii} = (h^i)^{-2}$, and the symbol $C_{ijk}$ is pairwise-symmetric in its indices with $C_{123}=1$ and zero otherwise. The $h^{i}$ fields are constrained by the relation $h^1 h^2 h^3 = 1$, which must be solved before taking variations of the action. Our conventions for Hodge duality as well as some useful expressions can be found in appendix \ref{subsec:Hodge}.

In the following we shall be concerned with solutions where the pseudoscalars $\chi^i$ are set to zero. This is not in general a consistent truncation, inasmuch as the pseudoscalar equations of motion then imply the constraints
\begin{align}\label{pseudoscalar constraint}
-f\,H_{ij}{\,\star_{4}F^{0}}\wedge F^{j} + \frac{1}{2}C_{ijk}\,F^{j}\wedge F^{k} = 0\,.
\end{align}
\noindent In order to fulfill these conditions we will consider solutions where $F^0$ is purely electric and the $F^i$ are purely magnetic. If we restrict to this case, we can write a simpler action from which we can derive the equations of motion, namely
\begin{align}\label{restricted STU model}
S
 =
-\frac{1}{2\kappa^{2}}\int d^{4}x\sqrt{|g|} \biggl[R  - \frac{e^{-\eta_0}}{4} F_{\mu\nu}^0 F^{0\,\mu\nu}- \frac{1}{2} \sum_{i=1}^3\biggl( \nabla_\mu \eta_i \nabla^\mu \eta_i
+ \frac{e^{2 \eta_i-\eta_{0}}}{2} F_{\mu\nu}^i  F^{i\,\mu\nu}\biggr)\biggr],
\end{align}
\noindent where $\kappa^{2} = 8\pi G_{4}$ ($\kappa$ has units of length), and we have introduced the shorthand notation
\begin{equation}
	\eta_{0} \equiv \eta_{1} + \eta_{2} + \eta_{3}\,.
\end{equation}
\noindent The scalar fields $\eta_{i}$ ($i=1,2,3$) are related to the scalars in \eqref{STUaction} through
\begin{align}\label{5d scalars}
h^i
&=
e^{\frac13 \eta_0 - \eta_i}\,,
\\
f
&=
e^{-\frac13 \eta_0}\,.
\end{align}
\noindent The corresponding equations of motion read\footnote{We note that the purely electric configurations of \cite{Cvetic:2011dn,Cvetic:2012tr} solve the equations of motion following from the action obtained from \eqref{restricted STU model} by dualizing the fields $F^{i}$ as $F^{i} \to -e^{\eta_0-2\eta_i}\left({\star_{4}F^i}\right)$.}
\begin{align}
0
={}&
\nabla_{\mu}\nabla^{\mu}\eta_{i} + \frac{1}{4}\Biggl[e^{-\eta_{0}}F^{0}_{\mu\nu}F^{0\mu\nu} + e^{-\eta_{0}}\sum_{j=1}^{3}\bigl(1-2\delta_{ij}\bigr) e^{2\eta_{j}}F_{\mu\nu}^{j} F^{j\mu\nu}\Biggr],
\\
0
={}&
\nabla_{\mu}\Bigl(e^{-\eta_{0}}F^{0\,\mu\nu}\Bigr),
\\
0
={}&
\nabla_{\mu}\Bigl(e^{-\eta_{0}+2\eta_{i}}F^{i\,\mu\nu}\Bigr),
\\
G_{\mu\nu}
={}&
\frac{1}{2}\Biggl[\sum_{i = 1}^{3}\left(\nabla_{\mu}\eta_{i}\nabla_{\nu}\eta_{i}-\frac{g_{\mu\nu}}{2} \nabla_\lambda \eta_i \nabla^\lambda \eta_i\right)
\nonumber\\
&
 + e^{-\eta_{0}}\left(F^{0\,\rho}_{\mu}F^{0}_{\nu\rho} - \frac{g_{\mu\nu}}{4}F^{0}_{\lambda\rho}F^{0\,\lambda\rho} \right)+ e^{-\eta_{0}}\sum_{i=1}^{3}e^{2\eta_{i}}\left(F^{i\,\rho}_{\mu}F^{i}_{\nu\rho} - \frac{g_{\mu\nu}}{4}F^{i}_{\lambda\rho}F^{i\,\lambda\rho} \right)\Biggr].
\end{align}

\subsection{Static ansatz}
In the present context we will be interested in static, spherically symmetric black hole backgrounds. As discussed above, in order to fulfill the constraint \eqref{pseudoscalar constraint} we furthermore consider an electric ansatz for $F^0$ and a magnetic ansatz for the $F^i$. Explicitly, our ansatz for the metric and matter fields reads
\begin{align}\label{4d ansatz 1}
ds_{4}^{2}
={}&
 -\frac{G(r)}{\sqrt{\Delta(r)}}dt^{2} + \sqrt{\Delta(r)}\left(\frac{dr^{2}}{X(r)} + d\theta^{2} + \frac{X(r)}{G(r)}\sin^{2}\theta\, d\phi^{2}\right)
 \\
 A^{0}
 ={}&
  A^{0}_{t}(r)\,dt
  \\
 A^{i} ={}&
 B_{i}\cos\theta\,d\phi
  \\
  \eta_{i}
  ={}&
   \eta_{i}(r)\, ,
   \label{4d ansatz 4}
\end{align}
\noindent where the constants $B^{i}$ ($i = 1,2,3$) are the magnetic charges. Einstein's equations are easily seen to imply $G(r) = \gamma X(r)$, where $\gamma = const$, and also $X''(r)=2$. Hence, without loss of generality we set
\begin{equation}
X(r)=G(r) =r^{2}-2mr\,.
\end{equation}
\noindent Given this ansatz, we first notice that the equation for $F^{0}$ implies
\begin{equation}\label{field strengths on shell}
	F^{0}_{rt} = q_{0}\frac{e^{\eta_{0}}}{\sqrt{\Delta}}\,,
\end{equation}
\noindent where the constant $q_{0}$ is the electric charge (up to normalization). The scalar equations then reduce to
\begin{align}
	\label{eqn:scalareom}
0
={}&
\Bigl(r(r-2m)\eta_{i}'\Bigr)'  -\frac{e^{\eta_{0}}}{2\sqrt{\Delta}} \left[q_{0}^{2}+ \sum_{j=1}^{3}\left(2\delta_{ij}-1\right)B_{j}^{2}\,e^{2\left(\eta_{j}-\eta_{0}\right)}\right].
\end{align}
\noindent Finally, one notices that the independent information contained in Einstein's equations amounts to one second order and one first order equation. These can be taken to be
\begin{align}
0
={}&
\frac{\Delta''}{\Delta} - \frac{3}{4}\left(\frac{\Delta'}{\Delta}\right)^{2} +\left( \eta_{1}'\right)^{2}+ \left(\eta_{2}'\right)^{2}+ \left(\eta_{3}'\right)^{2}
\\
0
={}&
\left(\frac{\Delta'}{2\Delta}\right)^{2} -\frac{2(r-m)}{r(r-2m)}\frac{\Delta'}{\Delta} +\frac{4}{r(r-2m)}
+\left( \eta_{1}'\right)^{2}+ \left(\eta_{2}'\right)^{2}+ \left(\eta_{3}'\right)^{2}
\nonumber\\
&
- \frac{e^{\eta_{0}}}{r(r-2m)\sqrt{\Delta}}\left[q_{0}^{2}+\sum_{i=1}^{3}e^{-2\left(\eta_{0}-\eta_{i}\right)}B_{i}^{2}\right].
\label{constraint}
\end{align}
\noindent The first of these equations is a linear combination of the $(t,t)$ and $(\phi,\phi)$ components of Einstein's equations, while the first order constraint is the $(r,r)$ component.

\subsection{Diagonalizing the non-linear system: decoupled modes and a family of static black hole solutions}\label{subsec: gen sol}
Quite remarkably, it is possible to diagonalize the full non-linear system of equations. To this end we introduce new fields $\phi_{0}$, $\phi_{i}$  defined as
\begin{align}
	\label{diagonal fields 1}
\phi_{0}(r) &=
\frac{1}{2}\log\left(\frac{\Delta(r)}{m^4}\right) - \eta_{1}(r)- \eta_{2}(r) -\eta_{3}(r)
 \\
 \phi_{1}(r) &=
   \frac{1}{2}\log\left(\frac{\Delta(r)}{m^4}\right) - \eta_{1}(r)+ \eta_{2}(r) + \eta_{3}(r)
\\
 \phi_{2}(r) &=
  \frac{1}{2}\log\left(\frac{\Delta(r)}{m^4}\right)+ \eta_{1}(r) - \eta_{2}(r) +\eta_{3}(r)
 \\
  \phi_{3}(r) &=
    \frac{1}{2}\log\left(\frac{\Delta(r)}{m^4}\right) +\eta_{1}(r) + \eta_{2}(r) -\eta_{3}(r)\,.
   \label{diagonal fields 2}
\end{align}
\noindent  Taking suitable linear combinations of the scalar and Einstein's equations one finds
\begin{align}\label{dec eqs 1}
0 ={}&
\Bigl(r\left(r-2m\right)\phi_{0}'(r)\Bigr)' + 2\left(\frac{q_{0}^{2}}{m^2}\,e^{-\phi_{0}(r)}-1\right)
\\
0={}&
\Bigl(r\left(r-2m\right)\phi_{i}'(r)\Bigr)' +2\left(\frac{B_{i}^{2}}{m^2}e^{-\phi_{i}(r)}-1\right)\,.
\label{dec eqs 2}
\end{align}
\noindent Upon solving these decoupled equations one has the solution for the original fields $\Delta(r)$ and $\eta_{i}(r)$, and the solution for $F^{0}$ is then given by \eqref{field strengths on shell}. Hence, we have effectively diagonalized the full non-linear system.

We have obtained general solutions to the decoupled equations \eqref{dec eqs 1}-\eqref{dec eqs 2}, each of which depends on two arbitrary integration constants. These generic solutions are not regular at the horizon $r=2m\,$, but upon imposing regularity they reduce to
\begin{align}\label{gen regular solution 1}
 \phi_{0}^{\text{reg}}(r)
 &=
  \log\left[\frac{q_{0}^{2}}{4m^{4}}\frac{\left(a_{0}^{2}\,r + 2m\right)^{2}}{1+a_{0}^{2}}\right]
 \\
  \phi_{j}^{\text{reg}}(r)
  &=
  \log\left[\frac{B_{j}^{2}}{4m^{4}}\frac{\left(a_{j}^{2}\,r + 2m\right)^{2}}{1+a_{j}^{2}}\right],
  \label{gen regular solution 2}
\end{align}
\noindent where the four independent constants $a_{0}$, $a_{i}$ parameterize a family of static black hole solutions. Close to the horizon, one finds
\begin{align}
\phi_{0}^{\text{reg}}(r \to 2m)
&=
 \log\Bigl(\frac{q_{0}^{2}}{m^2}\left(1+a_{0}^{2}\right)\Bigr) + \mathcal{O}\bigl(r-2m\bigr)\,,
 \\
 \phi_{j}^{\text{reg}}(r \to 2m)
 &=
  \log\Bigl(\frac{B_{j}^{2}}{m^2}\left(1+a_{j}^{2}\right)\Bigr) + \mathcal{O}\bigl(r-2m\bigr)\,.
\end{align}
\noindent Similarly, in the asymptotic region $r \to \infty$
\begin{align}
\phi_{0}^{\text{reg}}(r \to \infty)
&=
 \left\{
 \begin{array}{rrr}
 \log \frac{r^{2}}{m^2} + \mathcal{O}\left(1\right)\,,
  && a_{0} \neq 0 \\
\log \frac{q_{0}^{2}}{m^2}\,, && a_{0} =0
\end{array}
 \right.
 \\
 \phi_{j}^{\text{reg}}(r \to \infty)
 &=
   \left\{
 \begin{array}{rrr}
\log \frac{r^{2}}{m^2} + \mathcal{O}\left(1\right)\,,
  && a_{j} \neq 0 \\
\log \frac{B_{j}^{2}}{m^2}\,, && a_{j} =0
\end{array}
\right.
\end{align}
\noindent Going back to the original fields $\eta_{i}$ and $\Delta$, the solution reads
\begin{align}\label{general solution 1}
\Delta(r)
={}&
 \frac{\sqrt{q_{0}^{2}B_{1}^{2}B_{2}^{2}B_{3}^{2}}}{16m^{4}}\prod_{I=0}^{3}\frac{a_{I}^{2}\,r + 2m}{\sqrt{1+a_{I}^{2}}}
\\
e^{2\eta_{1}(r)}
={}&
\left|\frac{B_{2}B_{3}}{q_{0}B_{1}}\right|\sqrt{\frac{\left(1 +a_{0}^{2}\right)\left(1 +a_{1}^{2}\right)}{\left(1 +a_{2}^{2}\right)\left(1 +a_{3}^{2}\right)}}\frac{\left(a_{2}^{2}\,r + 2m\right)\left(a_{3}^{2}\,r + 2m\right)}{\left(a_{0}^{2}\,r + 2m\right)\left(a_{1}^{2}\,r + 2m\right)}
\\
e^{2\eta_{2}(r)}
={}&
\left|\frac{B_{1}B_{3}}{q_{0}B_{2}}\right|\sqrt{\frac{\left(1 +a_{0}^{2}\right)\left(1 +a_{2}^{2}\right)}{\left(1 +a_{1}^{2}\right)\left(1 +a_{3}^{2}\right)}}\frac{\left(a_{1}^{2}\,r + 2m\right)\left(a_{3}^{2}\,r + 2m\right)}{\left(a_{0}^{2}\,r + 2m\right)\left(a_{2}^{2}\,r + 2m\right)}
\\
e^{2\eta_{3}(r)}
={}&
\left|\frac{B_{1}B_{2}}{q_{0}B_{3}}\right|\sqrt{\frac{\left(1 +a_{0}^{2}\right)\left(1 +a_{3}^{2}\right)}{\left(1 +a_{1}^{2}\right)\left(1 +a_{2}^{2}\right)}}\frac{\left(a_{1}^{2}\,r + 2m\right)\left(a_{2}^{2}\,r + 2m\right)}{\left(a_{0}^{2}\,r + 2m\right)\left(a_{3}^{2}\,r + 2m\right)}\,.
\label{general solution 4}
\end{align}

It is worth emphasizing that, depending on how many of the constants $a_{0}$, $a_{i}\,$ are non-zero, the asymptotic behavior of $\Delta(r)$ in our family of solutions can be of the form $\Delta(r\to \infty)\sim  r^{\gamma}$, with $\gamma =0,1,\ldots, 4\,$. In particular, when $a_{0}=a_{i}=0$ (i.e. $\gamma = 0$) we obtain an asymptotically AdS$_{2}\times S^{2}$ black hole solution. When $\gamma \neq 0$, the metric displays a ``Lifshitz-covariance" of the form $t \to \lambda^{z}t\,$, $r\to\lambda^{2\theta/\gamma} r\,$, $ds^{2} \to \lambda^{\theta}ds^{2}\,$  in the $r \gg 2m$ region, where the dynamical exponent $z$ and the hyperscaling violation exponent\footnote{These metrics are in a sense a ``global" version of the planar black brane solutions that have been used to model condensed matter systems displaying hyperscaling violation; for some representative works, see \cite{Huijse:2011ef,Dong:2012se,Hartnoll:2012wm} and references therein.} $\theta$ are related by $\theta = \left(\frac{\gamma}{\gamma-2}\right)z$. We have checked explicitly that our family of solutions satisfies all the coupled equations of motion. In particular, the first order  constraint \eqref{constraint} is satisfied identically, and places no restriction on the values of the constants $a_{0}$, $a_{i}$.

\subsection{The ``original" and ``subtracted" geometries}\label{subsec:orig and subt}
The family of solutions found in section \ref{subsec: gen sol} contains as a particular case the solutions dubbed ``original" and ``subtracted" in \cite{Cvetic:2011hp,Cvetic:2011dn}.\footnote{As we have mentioned our solutions are related to the purely electric solutions of \cite{Cvetic:2011hp,Cvetic:2011dn} by the duality transformation $F^{i} \to -e^{\eta_0-2\eta_i}\left({\star_{4}F^i}\right)$. In particular the conformal factor $\Delta(r)$ and the scalars $\eta(r)$ are unaffected by this transformation, and it is in this sense that we use the same terminology to refer to the full solutions.} The original solution is given in terms of functions
\begin{equation}
p_{I}(r) = r + 2m\sinh^{2}\delta_{I}\,,
\end{equation}
\noindent ($I=0,1,2,3$) and it reads
\begin{align}
\Delta(r)
={}&
 \prod_{I=0}^{3}p_{I}(r)
 \\
 e^{-\eta_{i}(r)}
 ={}&
  p_{i}(r)\sqrt{\frac{p_{0}(r)}{p_{1}(r)p_{2}(r)p_{3}(r)}}
  \\
F_{rt}^{0}
={}&
 -m \frac{\sinh\left(2 \delta_{0}\right)}{p_{0}(r)^{2}}\,.
\end{align}
\noindent We then see that the original geometry is asymptotically flat in the $r\to\infty$ (i.e. $r \gg 2m$) region. Comparing with our general solution we can easily read off the electric and magnetic charges and the parameters $a_{I}$ in terms of the $\delta_{I}$:
\begin{gather}
q_{0}^{\text{orig}}  = -m \sinh(2\delta_{0})\,,\quad B_{i}^{\text{orig}}  = m\sinh\left(2\delta_{i}\right),
\\
a_{I}^{\text{orig}}  = \frac{1}{\sinh(\delta_{I})}\,.
\end{gather}

Similarly, the so-called subtracted geometry is given by
\begin{align}\label{subtracted solution}
\Delta(r)
={}&
  (2m)^{3}\Bigl[\left(\Pi_{c}^{2}-\Pi_{s}^{2}\right)r + 2m\,\Pi_{s}^{2}\Bigr]
 \\
 e^{\eta_{i}(r)}
 ={}&
\frac{1}{\sqrt{\Delta(r)}}\prod_{j \neq i}B_{j}
\\
F_{tr}^{0}
={}&
-\frac{16m^{4}}{\Delta^{2}(r)}\Pi_{c}\Pi_{s}B_{1}B_{2}B_{3}\,,
\label{subtracted solution 3}
\end{align}
\noindent  where the $B_{i}$ are the magnetic charges as before, and we can read off the electric charge as $q_{0} = -16m^{4}\left(B_{1}B_{2}B_{3}\right)^{-1}\Pi_{c}\Pi_{s}$.\footnote{Upon dualizing, we obtain a generalization of the solution presented in \cite{Cvetic:2011hp,Cvetic:2011dn} where we allow for a set of four independent $U(1)$ charges.} 
\noindent This solution is asymptotically conical for $r\rightarrow \infty$ \cite{Cvetic:2012tr}. Comparing with our general solution, we learn that the subtracted geometry has
\begin{equation}
a_{0}^{\text{subt}} = \sqrt{\frac{\Pi_{c}^{2}-\Pi_{s}^{2}}{\Pi_{s}^{2}}}\,,\qquad a^{\text{subt}}_{i}=0\,.
\end{equation}

As shown in \cite{Cvetic:2011hp}, the thermodynamics of the original and subtracted solutions matches if the parameters $\Pi_c$ and $\Pi_s$ are given as follows:
\begin{equation}
	\Pi_c = \prod_{I=0}^{3} \cosh \delta_I\,,\qquad	\Pi_s = \prod_{I=0}^{3} \sinh \delta_I\,.
\end{equation}

\begin{figure}[h]
\centering
\includegraphics[width=12cm]{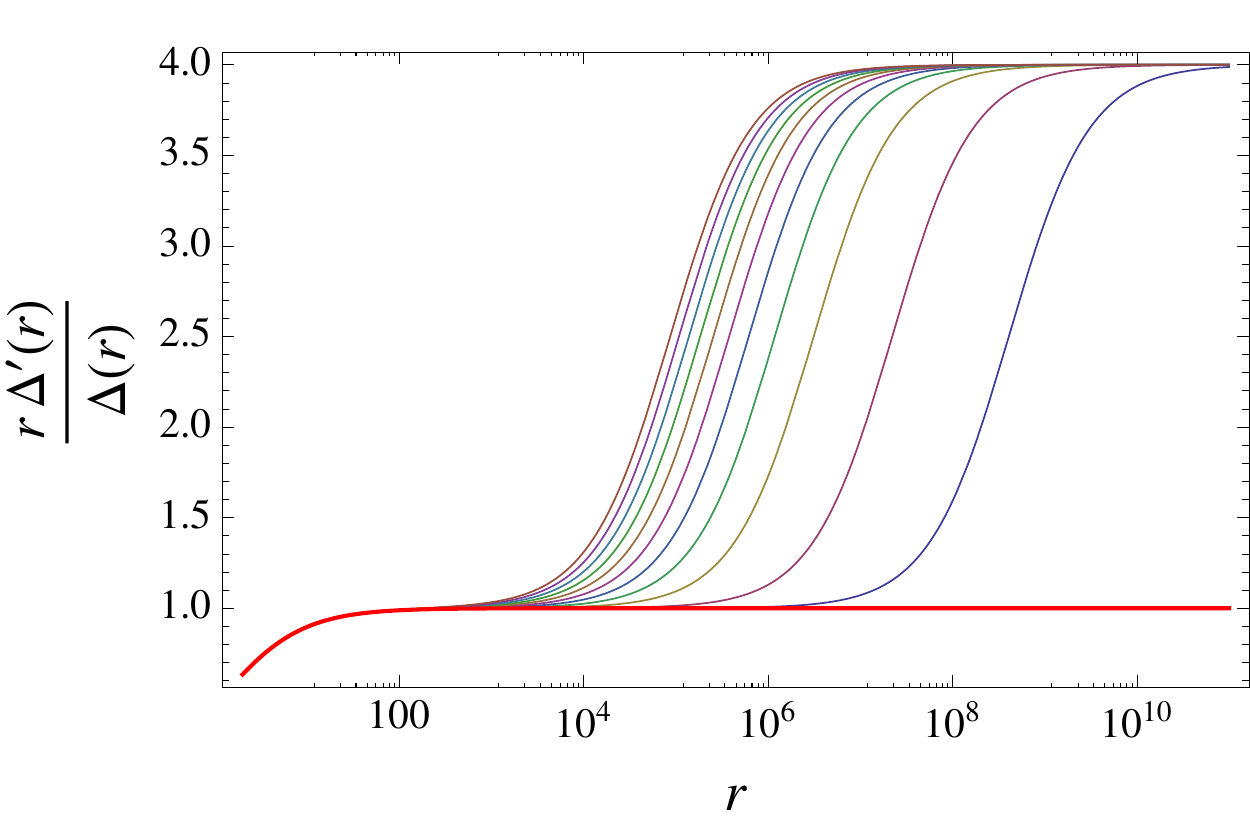}
\caption{Log plot of $\gamma(r) \equiv \frac{d\log(\Delta)}{d\log r}$ for the general solution \eqref{general solution 1}. The bottom red curve with $\gamma(r\gg 2m)=1$ corresponds to the subtracted geometry ($a_{1}=a_{2}=a_{3}=0$), while the various curves with $\gamma(r\gg 2m)=4$ correspond to the original geometry with different values for $a_{1}=a_{2}=a_{3}\equiv 1/\sinh(\delta)$. The different curves have increasingly larger values of $\delta$ towards the right, so we see that the original and subtracted geometries agree over a broader range in $r$ as the magnetic charges $B\sim \sinh(2\delta)$ increase.}
\label{fig:interpolating}
\end{figure}

As we indicated above, depending on how many of the parameters $a_{I}$ are zero the (large-$r$) asymptotic behavior of the conformal factor $\Delta(r)$ changes. While the asymptotically flat original geometry has all $a_{I} \neq 0$ and $\Delta_{\text{orig}} \sim r^{4}$ for large $r$ (i.e. $r\gg 2m$), we have shown that the subtracted solution has $a_{1}=a_{2}=a_{3}=0$ and therefore the conformal factor scales linearly $\Delta_{\text{subt}} \sim r$. Figure \ref{fig:interpolating} illustrates how we can smoothly interpolate between the subtracted and original geometries by dialing the parameters $a_{I}$. In particular notice that when all $\delta_i \gg 1$ a region emerges where the two solutions match to a very good approximation. It is in this sense that we refer to our family of solutions as an interpolating flow, with the different curves in figure \ref{fig:interpolating} corresponding to different points in the space of couplings on a putative dual field theory. In fact, in this limit (also known as dilute-gas approximation) one can think of the subtracted solution as coming from a decoupling limit of the original solution, as we will discuss in the next sections.

\section{Interpreting the flow between the original and subtracted geometries}\label{sec:flows}
In the previous section we described a four-parameter family of \textit{exact} static solutions of the STU model that interpolates between the original and subtracted geometries; depending on the choice of parameters, this family also includes geometries with different asymptotic behavior from that of the original geometry. We can thus view the subtracted geometry loosely as an IR endpoint of an RG flow starting from the original geometry.\\
 It is noteworthy that our solution implements explicitly the scaling limit discussed in \cite{Cvetic:2012tr} that extracts the subtracted solution from the original one. In the present section we will interpret this scaling limit as a flow between the original and subtracted geometries, while setting the stage for the AdS/CFT discussion to follow in section \ref{sec:uplift}.

Even though \eqref{general solution 1}-\eqref{general solution 4} (with $F^{0}$ given by \eqref{field strengths on shell}) is an exact solution of the full nonlinear equations, we find it instructive to discuss its linearized version. On the one hand the linearized analysis makes the discussion of regularity at the horizon cleaner, since this is related to a choice of state in the holographic context. Secondly, the sources that one gets by linearizing our family of exact solutions do not necessarily correspond to the sources of irrelevant perturbation theory, as we will discuss in detail below. Lastly, while in generic situations exact solutions to the nonlinear equations are not available, the principles behind the linearized analysis still apply. In particular, we will exhibit the existence of linearized modes of the subtracted geometry that start the flow to the original geometry when their sources are chosen correctly. As we will explicitly show in section \ref{sec:uplift}, upon uplifting the solutions to $5d$, these modes will turn out to be dual to irrelevant operators that deform the conformal field theory dual to the subtracted geometry.
\subsection{Linearized analysis}
We start our analysis by linearizing the field equations around the subtracted solution. Since the full non-linear equations of motion are diagonalized by the fields $\phi_I$ in \eqref{diagonal fields 1}-\eqref{diagonal fields 2}, we can simply consider
\begin{equation}
	\phi_I = \phi_I^{\text{subt}} + \delta\phi_I\,,
\end{equation}
where the $\delta\phi_I$ will be our linearized perturbations. The linearized equations are then
\begin{align}
	& r(r-2m) \delta\phi_0'' + 2(r-m) \delta\phi_0' - 2 \frac{q_0^2 B_1^2 B_2^2 B_3^2}{\Delta^2} \delta\phi_0 = 0\,,
	\\
	& r(r-2m) \delta\phi_i'' + 2(r-m) \delta\phi_i' - 2 \delta\phi_i = 0\,.
\end{align}
The equations for the $\delta\phi_i$'s are particularly simple, and we focus on those first. Changing to a new radial variable $x = \frac{r}{m}-1\,$, these equations take the form
\begin{equation}
	(1-x^2) \delta\phi_i''(x) - 2 x \delta\phi_i'(x) + 2 \delta\phi_i(x) = 0\,,
\end{equation}
which is a Legendre equation whose general solution is
\begin{equation}
	\delta\phi_i(x) = \alpha_i\, x + \beta_i \left(\frac{x}{2} \log\left(\frac{x+1}{x-1}\right) -1 \right).
\end{equation}
Of these two solutions, only one is regular at the horizon (which is located at $x=1$), therefore we must set $\beta_i=0$, or
\begin{equation}
\label{phiiperturbations}
	\delta\phi_i = \alpha_i \left(\frac{r}{m} - 1\right).
\end{equation}
Using the same variable $x$, and defining the parameters $b$ and $c$ as
\begin{equation}
	b = \frac{\Pi_c^2 - \Pi_s^2}{2 \sqrt2 \Pi_c \Pi_s}\,, \qquad c = \frac{\Pi_c^2 + \Pi_s^2}{2\sqrt2 \Pi_c \Pi_s}\,,
\end{equation}
the equation for $\delta\phi_0$ becomes
\begin{equation}
	(1-x^2) \delta\phi_0''(x) - 2 x \delta\phi_0'(x) + \frac{1}{(b x + c)^2} \delta\phi_0(x) = 0\,.
\end{equation}
The solution that is regular at the horizon in this case is given by
\begin{equation}
\label{phi0perturbations}
	\delta\phi_0 = \alpha_0 \frac{c x + b}{b x + c} = \alpha_0 \frac{(\Pi_c^2+\Pi_s^2) r - 2m \Pi_s^2}{(\Pi_c^2-\Pi_s^2) r + 2m \Pi_s^2}\,.
\end{equation}
Notice that the condition of regularity will translate into a functional relation between the normalizable and non-normalizable modes in the standard holographic setting.
\subsection{Perturbation theory and determination of the sources}
\label{sec:detsources}
In the previous section we showed that the solutions dubbed ``original'' and ``subtracted'' fit in a four-parameter family of solutions parametrized by $a_I$. In particular, recall that we have
\begin{equation}
	a^{\text{orig}}_i = \frac{1}{\sinh \delta_i}\,, \qquad a^{\text{subt}}_i = 0\,,
\end{equation}
while the two $a_0$'s are both different from zero. In order to go from the subtracted to the original geometry, we need to ``turn on'' the parameters $a_i$ and change the parameter $a_0$. We would like to understand this in terms of a flow that is started by linearized fluctuations around the subtracted background. This suggests that the sources $\alpha_I$ should be directly related to the parameters $a_I$. However, since we are turning on an irrelevant mode, at each order in perturbation theory higher powers of $r$ will be generated, therefore we need to treat the sources as infinitesimal quantities. It is easy to see that linearizing the general solution $\phi_i$ around $a_i^2 = 0$, at first order in $a_i^2$ one gets:
\begin{equation}
	\phi_i = \phi_i^{\text{subt}} + a_i^2 \left(\frac{r}{m} - 1\right) + \ldots\,,
\end{equation}
and we recognize the second term on the right-hand side as being the linearized perturbation of the previous subsection. Notice that the higher order terms do not contain terms linear in $r$, so the sources obtained by linearizing in $a_i^2$ are equivalent to the sources that one would obtain by extracting the coefficient of order $r$ in a power-series expansion. Therefore we should identify
\begin{equation}
	\label{eq:alphaiai1}
	\alpha_i = (a^{\text{orig}}_i)^2 = \frac{1}{\sinh^2\delta_i}\,.
\end{equation}
Analogously, we have
\begin{equation}
	\label{eq:alpha0a01}
	\phi_0 = \phi_0^{\text{subt}} + (a_0^2 - (a^{\text{subt}}_0)^2)\frac{\Pi_s^2}{\Pi_c^2} \frac{(\Pi_c^2+\Pi_s^2) r - 2m \Pi_s^2}{(\Pi_c^2-\Pi_s^2) r + 2m \Pi_s^2}\,,
\end{equation}
and as a consequence
\begin{equation}
	\alpha_0 =  \left((a^{\text{orig}}_0)^2 - (a^{\text{subt}}_0)^2\right) \frac{\Pi_s^2}{\Pi_c^2}\,.
\end{equation}
Notice however that the sources are in general \emph{not} infinitesimal. They do become infinitesimal in the limit where the three parameters $\delta_i$ become very large. In fact in this limit we obtain particularly simple expressions for the sources:
\begin{align}
	\alpha_i & \approx 4 \, e^{-2\delta_i}\,,
	\\
	\alpha_0 & \approx -4 \sum_i e^{-2\delta_i}\,.
\end{align}
Therefore, to leading order we have the relation:
\begin{equation}
	\frac{\delta \Delta}{\Delta} = \frac12 \sum_I \delta\phi_I = 0,
\end{equation}
that is, the metric is not changed to leading order in the parameters $e^{-2\delta_i}$.

Looking at the behavior of the linearized modes for very large $r$, one is led to the suspicion that the $\delta\phi_i$'s correspond to irrelevant perturbations while $\delta\phi_0$ seems to be associated to a marginal perturbation. In section \ref{sec:uplift} we will show that this suspicion is correct (after a suitable change of basis), and we will compute the quantum numbers of the operators in the dual CFT${}_2$ that we need to turn on to start the flow to the asymptotically flat original black hole. It is important to notice that these irrelevant perturbations do change the value of the matter fields in the interior, and in particular they are finite (i.e. non zero) at the horizon. This is in contrast to the extremal case, where irrelevant perturbations die off quickly in the interior and do not change the value of the fields at the horizon.

 Notice also that since we are turning on irrelevant deformations, there is no intrinsic (i.e. coordinate invariant) way to extract the sources for the dual operators. Their precise definitions must be supplemented with a perturbation scheme to compute higher order corrections. For example, it is easy to see that our choice for the $\alpha_i$'s is compatible with a scheme where the linear term in $r$ does not receive higher order corrections; however if we used a different radial coordinate, for example $r' = r + c$ where $c$ is a constant, this would not be true anymore. This ambiguity has an analog in quantum field theory, where the question of whether a source of an operator receives quantum corrections or not depends on the renormalization scheme. This ambiguity is obviously not present to leading order in perturbation theory. We will revisit this issue at the end of section \ref{sec:vacsoldualop}, where we will describe other possible choices for the sources.
 
\subsection{Range of validity of the linear approximation}
In many applications, one does not have the exact solutions, and often it is even impossible to solve the linearized equations exactly. In fact, in many interesting situations only the linearized modes in the asymptotic region are available, and a numerical treatment becomes necessary. It is therefore useful to investigate how one could approach this problem from a numerical perspective; we will then be able to compare the numerical results with the analytic results of the previous sections. The first step is to find a region that can be identified with the asymptotic region of the subtracted geometry (that is $\frac{r}{m} \gg 1$) but where the modes that start the flow to the original geometry are still small, so that they can be treated perturbatively. From the discussion in the previous sections, it is clear that this region should be
\begin{equation}
\label{eqn:window4d}
	1 < \frac{r}{m} \ll \frac{1}{\alpha}\,.
\end{equation}
where $\alpha$ is the smallest of the $\alpha_i$'s. Furthermore, we argued that the $\alpha_i$'s are related  to the parameters $\delta_i$, so that when the latter are large, $\alpha_i \approx 4\,e^{-2\delta_i}$. As anticipated in the previous section, this is when the three charges $B_1$, $B_2$, and $B_3$ are large compared to the fourth, $q_0\,$.

\begin{figure}[h!]
\centering
\includegraphics[width=10cm]{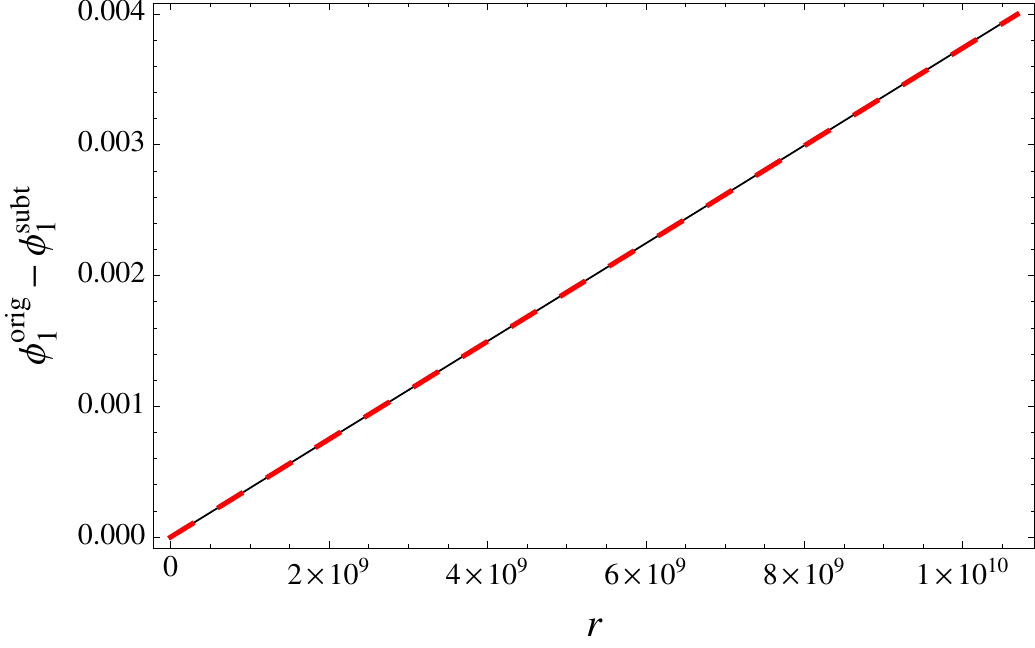}
\caption{The dashed red line represents the difference between the original and subtracted fields $\phi_1\,$. The solid black line is a linear function with slope $4e^{-2\delta_1}$. For this plot, we chose $\delta_0=\delta_1=\delta_2=\delta_3 = 15$, $m=1$, and the domain is $r \in [100,10^{-4} e^{2\delta_1}]\,$.}
\label{fig:phi1linear}
\end{figure}

Since we expect the difference between the original and subtracted solutions for the $\phi_i$'s to be linear in this intermediate region, it is possible to determine the sources for these three modes by means of a linear interpolation, as shown in figure \ref{fig:phi1linear}. The slope of the linear function turns out to be $4 e^{-2\delta_i}$, perfectly matching the results of the previous section.

\begin{figure}[h!]
\centering
\includegraphics[width=10cm]{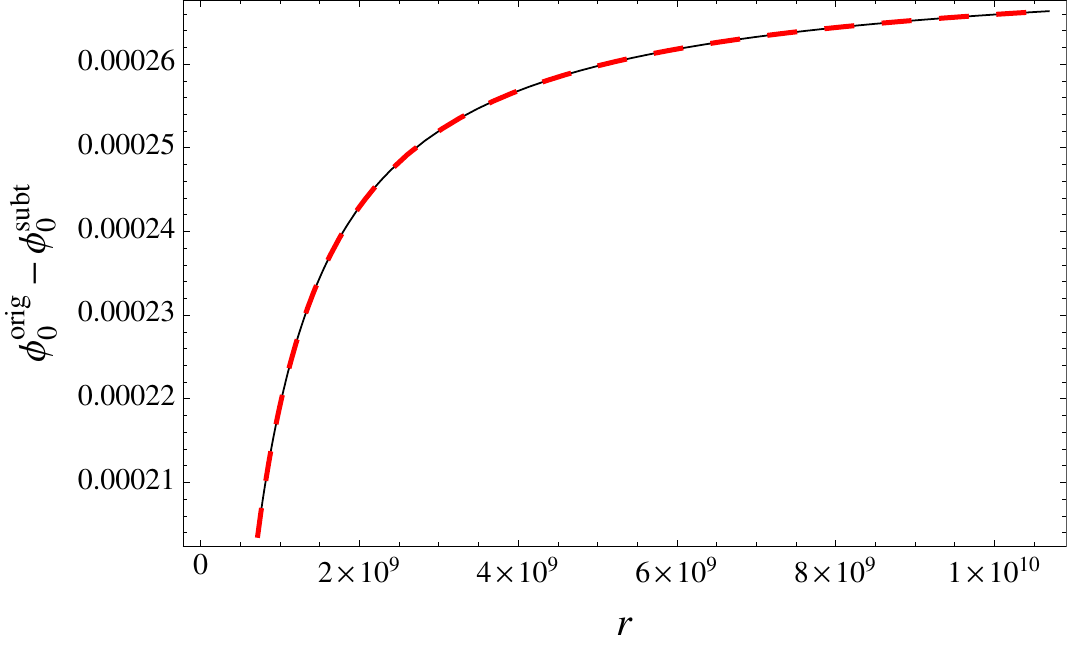}
\caption{The dashed red line represents the difference between the original and subtracted fields $\phi_0\,$. The solid black line is the function $4(e^{-2\delta_1} + e^{-2\delta_2} + e^{-2\delta_3}) \frac{(\Pi_c^2+\Pi_s^2) r - 2m \Pi_s^2}{(\Pi_c^2-\Pi_s^2) r + 2m \Pi_s^2}$. For this plot, we chose $\delta_0 = \delta_1=\delta_2=\delta_3 = 15$, $m=1$, and the domain is $r\in [100,10^{-4} e^{2\delta_1}]\,$.}
\label{fig:phi0linear}
\end{figure}

 We can also plot $\phi_0$ and the function $\alpha_0 \frac{(\Pi_c^2+\Pi_s^2) r - 2m \Pi_s^2}{(\Pi_c^2-\Pi_s^2) r + 2m \Pi_s^2}$. We see in figure \ref{fig:phi0linear} that the correct source for this mode is $\alpha_0 = -\alpha_1 - \alpha_2 - \alpha_3 = - 4(e^{-2\delta_1} + e^{-2\delta_2} + e^{-2\delta_3})$, confirming once again the analysis of the previous section. Before closing this section it is worth emphasizing that, by turning on different combinations of the sources, we can also flow to the various geometries with Lifshitz-like scaling discussed in section \ref{subsec: gen sol}.

\section{Uplifting and AdS/CFT interpretation} \label{sec:uplift}
In this section we uplift our $4d$ solutions to five dimensions, where an AdS/CFT interpretation of the flow is possible. As shown in \cite{Cvetic:2011hp}, the subtracted geometry uplifts to a BTZ black hole, which is asymptotically AdS$_3 \times S^2$. The linearized perturbations of the previous section uplift to linearized perturbations of the BTZ black hole, and we will explicitly show them to be dual to three irrelevant scalar operators with conformal weights $(h,\bar h) = (2,2)$. This allows us to give a more precise description of the dynamical realization of the conformal symmetry for the charged $4d$ black holes under study, while clarifying at the same time the limitations of this program.
\subsection{The $5d$ Lagrangian and equations of motion}
As shown in \cite{Virmani:2012kw}, the STU model \eqref{STUaction} can be obtained by dimensional reduction of the following $5d$ Lagrangian:
\begin{equation}
\label{eqn:Lagrangian5d}
\mathcal{L}_5 = R_5 {\,\star_5\mathds{1}}  - \frac{1}{2} H_{ij}{\,\star_5 dh^i }\wedge  dh^j
- \frac{1}{2}H_{ij}{\,\star_5 \tilde{F}^i} \wedge \tilde F^j + \frac{1}{6} C_{ijk}\, \tilde{F}^i \wedge \tilde{F}^j \wedge \tilde{A}^k,
\end{equation}
\noindent where $H_{ij}$ and $C_{ijk}$ are defined as in \eqref{STUaction}. The $4d$ and $5d$ line elements are related by
\begin{equation}\label{5d line element}
	ds_{5}^{2} = f^{-1} ds_{4}^{2}  + f^{2}\left(dz + A^{0}\right)^2\,,
\end{equation}
\noindent and the vector fields by
\begin{equation}
	\tilde A^i = \chi^i (dz + A^0) + A^i\,.
\end{equation}
\noindent The form of the $h^{i}$ scalars in our general four-parameter family of solutions is given in \eqref{general h1}-\eqref{general h3}. In particular, uplifting the subtracted solution \eqref{subtracted solution}-\eqref{subtracted solution 3} we discover that the $5d$ scalar fields are constant in this case:\footnote{Without loss of generality, in order to simplify the notation we assume that the magnetic charges satisfy $B_{i} > 0$ from now on. In the general case, the absolute value of various expressions involving products of the $B_{i}$ should be considered when appropriate.}
\begin{equation}\label{subtracted hs}
h_{\text{subt}}^{1} = \left(\frac{B_{1}^{2}}{B_{2}B_{3}}\right)^{1/3}\,,\qquad h_{\text{subt}}^{2} = \left(\frac{B_{2}^{2}}{B_{1}B_{3}}\right)^{1/3}
\,,\qquad  h_{\text{subt}}^{3} = \left(\frac{B_{3}^{2}}{B_{1}B_{2}}\right)^{1/3}\,.
\end{equation}
\noindent  As anticipated, the $5d$ subtracted geometry asymptotes to AdS$_{3}\times S^{2}$, and this will allow us to interpret the flow we found in the four-dimensional STU theory in terms of deformations of the CFT living on the boundary of the AdS$_{3}$ factor. The strategy we will follow consists of performing a consistent Kaluza-Klein reduction of the $5d$ theory on the two-sphere to obtain an effective $(2+1)$-dimensional theory. We will discover that the solutions of this effective theory with constant scalars correspond locally to AdS$_{3}\,$, and one of them uplifts precisely to the subtracted geometry in five dimensions. Linearizing the theory around this solution will then allow us to identify the dual operators associated with the flow between the original and subtracted geometries.

Before proceeding further we note that the model \eqref{eqn:Lagrangian5d} is slightly inconvenient in that the scalar fields satisfy the constraint $h^{1}h^{2}h^{3}=1$ which must be solved before taking variations of the action. Hence, we choose to work instead with unconstrained scalars $\Psi$ and $\Phi$ defined through \cite{Virmani:2012kw}
\begin{equation}\label{new scalar fields}
h^1 = e^{\sqrt{\frac{2}{3}} \Psi}, \quad h^2 = e^{-\frac{\Psi}{\sqrt{6}} -\frac{\Phi}{\sqrt{2}}}, \quad h^3 = e^{-\frac{\Psi}{\sqrt{6}} +\frac{\Phi}{\sqrt{2}}}\,,
\end{equation}
\noindent in terms of which
\begin{equation}
\label{eqn:Lagrangian5d b}
\mathcal{L}_5 = R_5 {\,\star_5\mathds{1}}  - \frac{1}{2} {\,\star_5 d\Psi }\wedge d\Psi -\frac{1}{2}{\,\star_5 d\Phi }\wedge  d\Phi
- \frac{1}{2}H_{ij}\left(\Psi,\Phi\right){\,\star_5 \tilde{F}^i} \wedge \tilde F^j + \frac{1}{6} C_{ijk}\, \tilde{F}^i \wedge \tilde{F}^j \wedge \tilde{A}^k\,.
\end{equation}
\noindent  The equations of motion for the matter fields are then
\begin{align}\label{5d eom}
0 ={}&
d\left(H_{ij}{\,\star_{5}\tilde{F}^{j}}\right) - \frac{C_{ijk}}{2}\tilde{F}^{j}\wedge\tilde{F}^{k}
\\
0={}&
d\bigl({\star_{5}d}\Psi\bigr) - \frac{1}{2}\frac{\delta H_{ij}}{\delta \Psi}{\,\star_5 \tilde{F}^{i}} \wedge \tilde{F}^{j}
\label{5d eom 2}
\\
0={}&
d\bigl({\star_{5}d}\Phi\bigr) - \frac{1}{2}\frac{\delta H_{ij}}{\delta \Phi}{\,\star_5 \tilde{F}^{i}} \wedge \tilde{F}^{j}\,.
\label{5d eom 3}
\end{align}
\noindent Similarly, Einstein's equations read
\begin{align}\label{5d Eins}
G_{\mu\nu} ={}&
\frac{1}{2}\biggl[\nabla_{\mu}\Psi\nabla_{\nu}\Psi - \frac{g_{\mu\nu}}{2}\nabla_{\lambda}\Psi\nabla^{\lambda}\Psi+\nabla_{\mu}\Phi\nabla_{\nu}\Phi - \frac{g_{\mu\nu}}{2}\nabla_{\lambda}\Phi\nabla^{\lambda}\Phi
\nonumber\\
&
\hphantom{\frac{1}{2}\biggl[}
 + H_{ij}\left(\tilde{F}^{i\hphantom{j}\rho}_{\mu}\tilde{F}^{j}_{\nu\rho} - \frac{g_{\mu\nu}}{4}\tilde{F}^{i}_{\lambda\rho}\tilde{F}^{j\,\lambda\rho}\right)\biggr]
\end{align}
\noindent and we recall that the only non-vanishing components of $H_{ij}\left(\Psi,\Phi\right)$ are given by $H_{ii}\left(\Psi,\Phi\right) = \left(h^{i}\left(\Psi,\Phi\right)\right)^{-2}\,$.

\subsection{Consistent Kaluza-Klein reduction}
The general structure of the uplifted line element \eqref{5d line element} is
\begin{align}
	ds_{5}^{2}
	={}&
	 e^{\frac{\eta_{0}}{3}}ds_{4}^{2}  + e^{-\frac{2\eta_{0}}{3}}\left(dz + A^{0}\right)^2
	 \label{uplifted line element}
	\\
	={}&
		 e^{\frac{\eta_{0}}{3}} \sqrt{\Delta(r)}\left(\frac{dr^{2}}{X(r)} -\frac{G(r)}{\Delta(r)}dt^{2} + \frac{e^{-\eta_{0}}}{\sqrt{\Delta(r)}}\left(dz + A^{0}\right)^2\right) +e^{\frac{\eta_{0}}{3}}\sqrt{\Delta(r)}\,ds^{2}\left(S^{2}\right)
\end{align}
\noindent where we assume that $A^{0}$ has no legs on the sphere directions (i.e. it is purely electric). It is easy to show that the subtracted geometry uplifts to a BTZ$\times S^2$ black hole. Nevertheless, it is more convenient to take a more general route that will allow us to characterize the general linear perturbations around the uplifted geometry. A  Kaluza-Klein (KK) Ansatz that includes all our uplifted solutions is
\begin{align}
ds_{5}^{2} ={}&
 ds^{2}_{\text{string}}(M) + e^{2U(x)}ds^{2}\left(Y\right)
 \\
 \tilde{F}^{i} ={}&
 -B_{i}\, \sin\theta\,d\theta\wedge d\phi
 \\
\Psi ={}&
 \Psi(x)
 \\
 \Phi = {}&
 \Phi(x)\,.
\end{align}
\noindent Here, $M$ is the $(2+1)$-dimensional ``external" manifold with coordinates $x = \{r,t,z\}$, and some metric that we keep arbitrary, and $Y$ is the ``internal" (compact) manifold, namely the two-sphere with radius $\ell_{S}$ and coordinates $y = \{\theta,\phi\}$. We pick the orientation such that the volume form on $Y$ is $\text{vol}_{2} = \ell_{S}^{2}\sin\theta\, d\theta\wedge d\phi\,$. The subscript ``string" in $ ds^{2}_{\text{string}}(M)$ is meant to remind us that the theory that will come out of the reduction will not be immediately in the $(2+1)$-dimensional Einstein frame, but rather in what could be called string frame. After performing the reduction we will translate the effective theory to Einstein frame before performing the AdS/CFT analysis. The radius of the sphere $\ell_S$ is set by the equations of motion to be:
\begin{equation}
	\ell_S = \left(B_1 B_2 B_3\right)^{1/3}\,.
\end{equation}
Notice however that we have the freedom to rescale $U$ in the reduction. The choice above guarantees that the radius of the reduced $3d$ (locally) AdS${}_3$ metric (in the final $3d$ Einstein frame) is equal to the radius of the (locally) AdS${}_3$ factor in the $5d$ geometry.

The details of the (consistent) KK reduction can be found in appendix \ref{subsec:KK}. Reducing the $5d$ equations of motion one finds that all reference to the two-sphere drops out, and the resulting three-dimensional equations of motion  \eqref{3d Psi eq string frame}, \eqref{3d Phi eq string frame}, \eqref{3d Einstein eq string frame} and \eqref{U equation string frame} follow from the effective (string frame) action
\begin{align}\label{effective action string frame}
S_{\text{string}} = -\frac{1}{16\pi G_3}\int d^{3}x \sqrt{|g|}\,e^{2U}\biggl[&
R + \frac{2}{\ell_{S}^{2}}e^{-2U}  - \frac{e^{-4U}}{2}\sum_{i=1}^{3}\frac{B_{i}^{2}}{\ell_{S}^{4}}H_{ii}(\Psi,\Phi)
\nonumber\\
&
+ 2\left(\nabla U\right)^{2} -\frac{1}{2}\left(\nabla\Psi\right)^{2}-\frac{1}{2}\left(\nabla\Phi\right)^{2}\biggr].
\end{align}
\noindent The three dimensional Newton's constant $G_3$ is fixed in terms of the normalization of the $5d$ action and the volume of the internal manifold, which are in turn related to the $4d$ Newton's constant $G_4$:
\begin{equation}
	G_3 = \frac{1}{4\pi\ell_S^2}G_5 = \frac{R_{z}}{2\ell_S^2}G_4\,.
\end{equation}
\noindent Here, $R_{z}$ is the radius of the circle on which we reduce to go from the $5d$ theory \eqref{eqn:Lagrangian5d} to the $4d$ STU model, and is in principle arbitrary. The next step consists in passing to three-dimensional Einstein frame by performing a Weyl rescaling of the metric on $M$. Denoting with a subscript $(E)$ the quantities in Einstein frame, the transformation we need is
\begin{equation}
ds^{2}_{\text{string}}(M) = e^{-4U}ds^{2}_{\text{(E)}}(M)\,,
\end{equation}
\noindent which in particular implies
\begin{equation}
R = e^{4U}\left[R_{(E)} + 8\square_{(E)}U - 8\left(\nabla_{(E)} U\right)^2\right].
\end{equation}
\noindent  It follows that the Einstein frame effective action is (after dropping a surface term)
\begin{align}\label{effective action Einstein frame}
S_{\text{(E)}} = -\frac{1}{16 \pi G_3}\int d^{3}x \sqrt{|g_{\text{(E)}}|}\,\biggl[&
R_{(E)} - 6\left(\nabla_{(E)} U\right)^{2} -\frac{1}{2}\left(\nabla_{(E)}\Psi\right)^{2}-\frac{1}{2}\left(\nabla_{(E)}\Phi\right)^{2}
\nonumber\\
&
 + \frac{2}{\ell_{S}^{2}}e^{-6U}- \frac{e^{-8U}}{2}\sum_{i=1}^{3}\frac{B_{i}^{2}}{\ell_{S}^{4}}H_{ii}(\Psi,\Phi)\biggr].
\end{align}
\noindent In a slight abuse of notation, we will drop the subscript $(E)$ from now on because we will be working exclusively in Einstein frame. The equations of motion are then
\begin{align}\label{Einstein frame eq 1}
G_{\mu\nu}
={}&
-\frac{g_{\mu\nu}}{\ell_{S}^{2}}\left[-e^{-6U} + \frac{e^{-8U}}{4}\sum_{i=1}^{3}\frac{B_{i}^{2}}{\ell_{S}^{2}}H_{ii}(\Psi,\Phi)\right]+
\frac{1}{2}\tilde{T}_{\mu\nu}
\\
0={}&
\nabla_{\mu}\nabla^{\mu}\Psi - \frac{e^{-8U}}{2}\sum_{i=1}^{3}\frac{B_{i}^{2}}{\ell_{S}^{4}}\frac{\delta H_{ii}(\Psi,\Phi)}{\delta\Psi}
\\
0={}&
\nabla_{\mu}\nabla^{\mu}\Phi - \frac{e^{-8U}}{2}\sum_{i=1}^{3}\frac{B_{i}^{2}}{\ell_{S}^{4}}\frac{\delta H_{ii}(\Psi,\Phi)}{\delta\Phi}
\\
0={}&
\nabla_{\mu}\nabla^{\mu}U - \frac{e^{-6U}}{\ell_{S}^{2}} + \frac{e^{-8U}}{3}\sum_{i=1}^{3}\frac{B_{i}^{2}}{\ell_{S}^{4}}H_{ii}(\Psi,\Phi)\, ,
\label{Einstein frame eq 4}
\end{align}
\noindent where we defined the ``kinetic" part of the stress tensor as
\begin{equation}
\tilde{T}_{\mu\nu}= \nabla_{\mu}\Psi\nabla_{\nu}\Psi - \frac{g_{\mu\nu}}{2}\left(\nabla \Psi\right)^{2}+\nabla_{\mu}\Phi\nabla_{\nu}\Phi - \frac{g_{\mu\nu}}{2}\left(\nabla \Phi\right)^{2} +12 \nabla_{\mu}U\nabla_{\nu}U - 6g_{\mu\nu}\left(\nabla U\right)^{2}\,.
\end{equation}

\subsection{Asymptotically AdS$_{3}$ solutions and dual operators}\label{sec:vacsoldualop}
We will consider solutions where the scalars take constant values $U = \bar{U}$, $\Psi = \bar{\Psi}$, $\Phi = \bar{\Phi}$, so that $\tilde{T}_{\mu\nu}=0$. In such a background, the equations \eqref{Einstein frame eq 1}-\eqref{Einstein frame eq 4} reduce to
\begin{align}
G_{\mu\nu}
={}&
-\Lambda_{\text{\scriptsize{eff}}}\, g_{\mu\nu}
\label{vacuum Einstein}
\\
0={}&
\sum_{i=1}^{3}B_{i}^{2}\left.\frac{\delta H_{ii}(\Psi,\Phi)}{\delta\Psi}\right|_{\bar{\Psi},\bar{\Phi}}
 \label{vacuum Psi}
\\
0={}&
\sum_{i=1}^{3}B_{i}^{2}\left.\frac{\delta H_{ii}(\Psi,\Phi)}{\delta\Phi}\right|_{\bar{\Psi},\bar{\Phi}}
 \label{vacuum Phi}
\\
e^{2\bar{U}} ={}&
 \frac{1}{3}\sum_{i=1}^{3}\frac{B_{i}^{2}}{\ell_{S}^{2}}H_{ii}(\bar{\Psi},\bar{\Phi})\,,
 \label{vacuum U}
\end{align}
\noindent where the effective cosmological constant $\Lambda_{\text{\scriptsize{eff}}}$ is given by
\begin{equation}
\Lambda_{\text{\scriptsize{eff}}}= \frac{1}{\ell_{S}^{2}}\left[-e^{-6\bar{U}} + \frac{e^{-8\bar{U}}}{4}\sum_{i=1}^{3}\frac{B_{i}^{2}}{\ell_{S}^{2}}H_{ii}(\bar{\Psi},\bar{\Phi})\right] =  -\frac{e^{-6\bar{U}}}{4\ell_{S}^{2}} \,,
\end{equation}
\noindent and we used \eqref{vacuum U} in the last equality. In three dimensions, the only solutions to Einstein's equations with negative cosmological constant are locally AdS${}_{3}$; the effective AdS$_{3}$ length $L$ in our case is then given by
\begin{equation}
L^{2} = -\frac{1}{\Lambda_{\text{\scriptsize{eff}}}} =  4\,e^{6\bar{U}}\ell_{S}^{2}\,.
\end{equation}
\noindent There is in fact a unique solution to equations \eqref{vacuum Psi}-\eqref{vacuum U} for the scalars, given by
\begin{equation}
e^{\bar{U}} =  \left(\frac{B_{1}B_{2}B_{3}}{\ell_{S}^{3}}\right)^{1/3} = 1\,,\qquad e^{\bar{\Psi}} = \left(\frac{B_{1}^{2}}{B_{2}B_{3}}\right)^{\frac{1}{\sqrt{6}}}\,,\qquad e^{\bar{\Phi}} = \left(\frac{B_{3}}{B_{2}}\right)^{\frac{1}{\sqrt{2}}}\,.
\end{equation}
\noindent Notice in particular that $\bar U = 0$. Comparing with \eqref{subtracted hs}-\eqref{new scalar fields}, we see that these are precisely the values corresponding to the subtracted geometry.\footnote{For completeness, the explicit form of the $3d$ scalars in our general family of solutions is given in \eqref{general U}-\eqref{general Phi}.} Moreover, as it follows from \eqref{vacuum Einstein}, the metric of this three-dimensional solution is locally AdS$_{3}$ with radius
\begin{equation}
L = 2\,e^{3\bar{U}}\ell_{S} = 2\ell_{S}= 2 \left(B_{1}B_{2}B_{3}\right)^{1/3}\,.
\end{equation}
We will describe the global properties of the solution that corresponds to the subtracted geometry in the following subsection.

We can determine the operator content of the dual field theory from the action \eqref{effective action Einstein frame}: we have the stress tensor coupling to the massless graviton, and three scalar operators that couple to the boundary values of $U$, $\Psi$, $\Phi$. Following the standard AdS/CFT dictionary, in order to compute the conformal dimensions of these operators we need to obtain the masses of the linearized bulk fields around the solution corresponding to the subtracted geometry. Linearizing the equations we find that the fluctuations of the three bulk scalars decouple and in fact satisfy the same equation:
\begin{align}
0
={}&
 \nabla_{\mu}\nabla^{\mu}\delta F -\frac{8}{L^{2}}\,\delta F\,,
\end{align}
\noindent where $\delta F$ stands for any of $\delta U$, $\delta \Psi$, $\delta \Phi$. Therefore, the masses are given by
\begin{equation}
m_{\delta U}^{2} = m_{\delta \Psi}^{2} = m_{\delta \Phi}^{2} =  \frac{8}{L^{2}}\,,
\end{equation}
\noindent and according to the standard dictionary we conclude that the three scalar operators in the dual theory are irrelevant, with conformal dimension $\Delta =4$\,.

\subsection{Irrelevant deformation of the CFT}
Finally, we relate the $4d$ modes of section \ref{sec:flows}, parametrized by the $\alpha_I$'s, to the linearized modes of the $3d$ theory. The scalars are
\begin{align}
	\delta U & = \frac{1}{6m} \left(\alpha_1 + \alpha_2 + \alpha_3\right) (r-m)\,
	\\
	\delta \Psi & = \frac{1}{2\sqrt6 m} \left(2\alpha_1 - \alpha_2 - \alpha_3\right) (r-m)\,
	\\
	\delta \Phi & = \frac{1}{2\sqrt2 m} \left(\alpha_3 - \alpha_2\right)(r-m)\,,
\end{align}
\noindent corresponding to non-normalizable modes. To identify the marginal mode, a little work is required. As explained in \cite{Cvetic:2011dn}, the uplifted subtracted geometry can be cast in the BTZ form with the change of coordinates (we work in a gauge where $A_0 \to 0$ as $r \to \infty$):
\begin{align}
	\rho^2 &= \frac{R_z^2}{\ell_S^4} \Delta(r) = \frac{R_z^2}{\ell_S^4} (2m)^3\Bigl( \left(\Pi_c^2 - \Pi_s^2\right)r + 2m \Pi_s^2\Bigr)\\
	t &= \frac{R_z}{2\ell_S^4} (2m)^3 \left(\Pi_c^2 - \Pi_s^2\right) t_3\\
	z &= - R_z \phi_3\,,
\end{align}
so that the metric reads
\begin{equation} \label{BTZ}
	ds^2 = -\frac{ (\rho^2 - \rho_+^2)(\rho^2 - \rho_-^2)}{L^2 \rho^2} dt_3^2 + \frac{L^2 \rho^2}{(\rho^2 - \rho_+^2)(\rho^2 - \rho_-^2)} d\rho^2 + \rho^2 \left(d\phi_3 + \frac{\rho_+ \rho_-}{L \rho^2} dt_3\right)^2 ,
\end{equation}
with the position of the inner ($\rho_{-}$) and outer ($\rho_{+}$) horizons given by
\begin{equation} \rho_+ = \frac{16 m^2 R_z}{L^2}\Pi_c\,,\qquad
 \rho_- = \frac{16 m^2 R_z}{L^2} \Pi_s\,.
\end{equation}
\noindent The left- and right-moving temperatures are then
\begin{equation}
T_{L} = \frac{\rho_{+} + \rho_{-}}{2\pi L^{2}} =  \frac{8 m^2 R_{z}}{\pi L^4} \left(\Pi_c + \Pi_s\right)\,,\qquad
T_{R} = \frac{\rho_{+} - \rho_{-}}{2\pi L^{2}} =\frac{8 m^2 R_{z}}{\pi L^4}\left(\Pi_c - \Pi_s\right),
\end{equation}
\noindent and the black hole mass, angular momentum, entropy density and temperature are
\begin{align}
	M &= \frac{1}{8G_{3}}\left( \frac{\rho_{+}^{2} + \rho_{-}^{2}}{L^{2}}\right) =  \frac{32 m^4 R_{z}^2}{L^6 G_3}\left(\Pi_c^2 + \Pi_s^2\right)
	\\
	J &= \frac{1}{8G_{3}}\left(\frac{2\rho_{+}\rho_{-}}{L}\right)=\frac{64 m^4 R_{z}^2 }{L^5 G_3}\Pi_c \Pi_s
	\\
    S & = \frac{\left(4\pi \rho_{+}\right)}{8G_{3}} = \frac{8\pi m^{2}R_{z}}{L^{2}G_{3}}\Pi_{c}
    \\
	T &= \frac{2T_{L}T_{R}}{T_{L} + T_{R}}=\frac{8 m^2 R_{z}}{\pi L^4}\left(\frac{\Pi_c^2 - \Pi_s^2}{ \Pi_c}\right)\,.
\end{align}
When we uplift the perturbations (\ref{phiiperturbations}) and (\ref{phi0perturbations}), we find that they superficially destroy the BTZ asymptotics. This is due to the fact that all the independent perturbations in the $4d$ theory involve a change in the metric. However, since the $3d$ linearized Einstein's equations decoupled from the matter fields, the solution must still be locally AdS$_3$ and the uplifted perturbations must correspond to a change in the BTZ parameters up to diffeomorphisms. Indeed, we can perform a linearized diffeomorphism
\begin{equation}
	\delta g_{\mu\nu} = 2\nabla_{(\mu}\xi_{\nu)}\,,
\end{equation}
that brings the metric to the original BTZ form, with the change of parameters:
\begin{align}
	\delta \rho_+ &= \Bigl(\alpha_0+\sum_i\alpha_i\Bigr)\frac{4m^2 R_z}{L^2} \Pi_c\\
	\delta \rho_- &=  -\Bigl(\alpha_0+\sum_i\alpha_i\Bigr)\frac{4m^2 R_z}{L^2} \Pi_s\,.
\end{align}
Incidentally, this shows that the marginal mode is non-normalizable from the AdS$_{3}$ perspective, and that there is a non-trivial change of basis between the independent $4d$ modes and the $3d$ modes. We can translate this into a change of mass and angular momentum of the BTZ black hole:
\begin{align}
	\delta M &= \delta\left( \frac{\rho_+^2 + \rho_-^2}{8G_3 L^2}\right) = \Bigl(\alpha_0+\sum_i\alpha_i\Bigr)\frac{16 m^4 R_z^2}{G_3 L^6}\left(\Pi_c^2-\Pi_s^2\right)\\
	\delta J &= 0\,.
\end{align}
Notice that the variations of the physical BTZ parameters vanish when
\begin{equation}
	\alpha_0 = - \sum_i \alpha_i\,.
\end{equation}
As we now explain, we can choose our sources so that they satisfy the relation above, and this corresponds to a scheme where the entropy of the black hole does not change order by order in perturbation theory.

Recall that the precise relation between the parameters $\alpha_I$ and the parameters $a_I$ that describe the family of exact black hole solutions depends on the renormalization scheme, as explained at the end of section \ref{sec:detsources}. The choice \eqref{eq:alphaiai1}-\eqref{eq:alpha0a01} is one possibility, but here we will present an alternative that is more natural from the point of view of AdS/CFT. In quantum field theory one has the freedom to redefine the sources at each order in perturbation theory, so that
\begin{equation}
	J = J_0 + \lambda J_1 + \ldots + \lambda^n J_n + \ldots\,,
\end{equation}
where $\lambda$ is the coupling constant. It is customary to choose a scheme where
\begin{equation}
	J = J_0\,,
\end{equation}
i.e. where the source is not renormalized. From the point of view of AdS/CFT, this means that the coefficient of $\rho^{\Delta-d}$ that corresponds to the source of the dual operator does not change at higher order in perturbation theory. This corresponds to the choice
\begin{equation}
	\alpha_i = \frac{\Pi_c^2 - \Pi_s^2}{\Pi_c^2\,\sinh^2\delta_i - \Pi_s^2\,\cosh^2\delta_i} \approx 4 e^{-2\delta_i}\,,
\end{equation}
showing once again that the leading contribution is independent of the scheme. It is possible to do the same for the metric mode associated to the dual stress tensor, but in this context it seems more natural to choose a scheme where the $4d$ metric does not change at the horizon order by order in perturbation theory. This yields
\begin{equation}
	\alpha_0 = - \sum_{i=1}^{3} \alpha_i \approx -4\sum_{i=1}^{3} e^{-2\delta_i}\,,
\end{equation}
which once again agrees with the previous results to leading order in $e^{-2\delta_i}\,$. Notice that this choice corresponds to keeping the physical parameters of the BTZ black hole fixed, so that at first order the marginal mode associated to the metric is turned off. We conclude that the flow to the original geometry is started by turning on three irrelevant operators in the dual CFT.

\subsection{Irrelevant mass scale and range of validity of the CFT description}
Finally, we can determine the mass scale set by the irrelevant deformations, which represents the UV cutoff of the dual field theory. Consider the asymptotic behavior of the field $\delta U$:
\begin{equation}
	\delta U = \frac{\ell_S^4 \sum_{i=1}^{3} \alpha_i}{3 R_{z}^2 (2m)^4 \left(\Pi_c^2 - \Pi_s^2\right)}\rho^2 + \ldots\,.
\end{equation}
As a consequence, the source of the operator dual to $U$ reads
\begin{equation}
\label{eqn:source}
	J_U =  \frac{L^8 \sum_{i=1}^{3} \alpha_i}{48 R_{z}^2 (2m)^4 \left(\Pi_c^2 - \Pi_s^2\right)} = \frac{1}{12\pi^2 T_L T_R} \sum_i \alpha_i\,.
\end{equation}

Recall that the temperature in a CFT sets an infrared cutoff, while the mass scale of the irrelevant deformation sets an ultraviolet cutoff. Equation \eqref{eqn:source} shows that when the $\alpha_i$'s are of order 1, the infrared cutoff and the ultraviolet cutoff are of the same order, so that there is no regime where the conformal field theory description is meaningful. On the other hand, when the $\alpha_i$'s become small (or $\delta_i \gg 1$), an energy window appears where perturbation theory on the CFT should be a good description of the system:
\begin{equation}
	1 < \frac{E^2}{T_L T_R} \ll \frac{1}{\alpha}\,.
\end{equation}
This is the CFT analog of the condition \eqref{eqn:window4d} that we have identified in the $4d$ system.

We can phrase the result above in terms of standard effective field theory. The contributions of irrelevant couplings to a process characterized by an energy scale $E$ are typically suppressed by powers of $E/M$, where $M$ is the UV cutoff set by the irrelevant couplings. In our case we have
\begin{equation}
	M^2 \approx \frac{1}{\alpha} T_L T_R\,,
\end{equation}
and this should be compared to the IR cutoff of the system, that is the temperature. One way to see this is that contributions from the region $E\sim M$ to thermal expectation values are suppressed by a factor $e^{-\beta M}$; since these contributions cannot be reliably computed in effective field theory, we require $\beta M \gg 1$. In this sense, $M$ is very large when the $\alpha$'s are small, opening up a range of energies where effective field theory becomes meaningful. It is precisely in this region that CFT (plus perturbation theory) becomes a good description of the system.

\section{Discussion}\label{sec:discussion}
It has been recently argued that certain questions involving the entropy and thermodynamics of four-dimensional asymptotically flat non-extremal black holes can be elucidated by replacing the original geometry by one with a different conformal factor, dubbed subtracted geometry. The replacement modifies the asymptotics while preserving the near-horizon behavior of the original black hole, in such a way that the role of an underlying conformal symmetry becomes manifest, shedding light on the form of the entropy for black holes away from extremality \cite{Cvetic:2011hp,Cvetic:2011dn,Cvetic:2012tr}. Building on these works, we have shown that four-dimensional, static, asymptotically flat non-extremal black holes with one electric and three magnetic charges can be connected to their corresponding subtracted geometry by a flow which we have constructed explicitly in the form of an interpolating family of solutions. Upon uplifting the construction to five dimensions the subtracted geometry asymptotes to AdS$_{3}\times S^{2}$, and an AdS/CFT interpretation of the flow is readily available as the effect of irrelevant perturbations in the conformal field theory dual to the AdS$_{3}$ factor. In particular, we have identified the quantum numbers of the deformations responsible for the flow and showed that they correspond to three scalar operators with conformal weights $(h,\bar{h}) = (2,2)$.

As discussed in detail in section \ref{sec:flows} and \ref{sec:uplift}, the mass scale associated to such irrelevant perturbations becomes very large compared to the temperature when the magnetic charges are large. In this limit, it is reasonable to expect that some dynamical questions can be approximately answered by means of perturbation theory in the CFT${}_2\,$. At least in the static limit, our construction then puts the procedure followed in \cite{Cvetic:2011hp,Cvetic:2011dn} on a somewhat more concrete footing. On the other hand, away from this limit the ultraviolet cutoff set by the irrelevant deformations becomes of the same order of the infrared cutoff set by the temperature, and the dual CFT captures an increasingly smaller subset of the dynamics, making the usefulness of such an approach doubtful.

It would be of interest to extend our analysis to include rotating four-dimensional black holes. Even though we do not expect any conceptual difficulties, the rotating case is technically more challenging: in the $5d$ uplifted geometry the two-sphere $S^2$ is fibered non-trivially over AdS${}_3\,$, and the modes that start the flow presumably involve non-trivial harmonics on the sphere. This case will be addressed elsewhere. It would also be very interesting to set up the perturbation scheme in the dual CFT${}_2$ and determine what observables of the $4d$ black hole can be reliably computed in terms of perturbation theory in the irrelevant couplings. Effective field theory makes sense only up to the scale set by the irrelevant deformations. However, our perturbations can be resummed geometrically to all orders, allowing us to go beyond the region where perturbation theory is meaningful and reach the asymptotically flat region. From the field theoretic perspective, it is then natural to wonder whether this allows us to say something about the regime where effective field theory breaks down. Similar questions can be considered for the black holes with Lifshitz-like asymptotics that can be obtained by turning on only a subset of the irrelevant deformations.

\acknowledgments
It is a pleasure to thank Borun Chowdhury, Geoffrey Comp\`ere and Balt van Rees for helpful conversations, and especially Finn Larsen for discussions that helped to motivate this project. We also thank Natalia Pinzani-Fokeeva for collaboration on an early stage of this project. This work is part of the research programme of the Foundation for Fundamental Research on Matter (FOM), which is part of the Netherlands Organization for Scientific Research (NWO).

\appendix
\section{Conventions and useful formulae}
\subsection{Hodge duality \label{subsec:Hodge}}
Let $\omega$ be a $p$-form in $D$-dimensions,
\begin{equation}
\omega = \frac{1}{p!}\omega_{\mu_{1}\ldots \mu_{p}}\,dx^{\mu_{1}}\wedge\ldots\wedge dx^{\mu_{p}}\,.
\end{equation}
\noindent We define the action of the Hodge star on the basis of forms as
\begin{equation}
\star(dx^{\mu_1}\wedge \cdots \wedge dx^{\mu_p}) = \frac{1}{(D-p)!}{\epsilon_{\nu_{1} \ldots \nu_{D-p}}}^{\mu_1\ldots \mu_p}\, dx^{\nu_{1}}\wedge \cdots \wedge dx^{\nu_{D-p}}\, ,
\end{equation}
\noindent where $\epsilon_{\mu_{1}\ldots \mu_{D}}$ are the components of the Levi-Civita tensor. Equivalently, in components we find
\begin{equation}
 \left(\star\,\omega\right)_{\mu_{1}\ldots \mu_{D-p}} = \frac{1}{p!}\epsilon_{\mu_{1}\ldots \mu_{D-p}\,\nu_{1}\ldots \nu_{p}}\, \omega^{\nu_{1}\ldots \nu_{p}}\, .
\end{equation}
\noindent  If $\varepsilon_{\mu_{1}\ldots \mu_{D}}$ denotes the components of the Levi-Civita \textit{symbol} (a tensor density), we have
\begin{equation}
\epsilon_{\mu_{1}\ldots \mu_{D}} = \sqrt{|g|}\,\varepsilon_{\mu_{1}\ldots \mu_{D}} \quad \Leftrightarrow\quad \epsilon^{\mu_{1}\ldots \mu_{D}} = \frac{(-1)^{t}}{\sqrt{|g|}}\varepsilon^{\mu_{1}\ldots \mu_{D}}
\end{equation}
\noindent where $t$ denotes the number of timelike directions, and we have adopted the convention that the Levi-Civita symbol $\varepsilon$ with up or down indices is the same. The volume element is given by
\begin{equation}
\star\mathds{1} = \sqrt{|g|}\,d^{D}x \equiv \mbox{vol}_{D} \qquad \Rightarrow \qquad \star\mbox{vol}_{D} =  (-1)^{t} \mathds{1}\,.
\end{equation}
\noindent A useful observation is that, for any two $p$-forms $A$ and $B$,
\begin{align}\label{square of forms}
\star A \wedge B = \star B\wedge A
={}&
\frac{1}{p!}\,A^{\mu_{1} \ldots \mu_{p}}B_{\mu_{1}\ldots \mu_{p}}\, \text{vol}_{D}\,.
\end{align}
\noindent  Similarly, if $\phi$ is a scalar it follows
\begin{equation}
d{\star d\phi} = \left(-1\right)^{D-1}\nabla_{\mu}\nabla^{\mu}\phi\,\text{vol}_{D} \qquad\Rightarrow\qquad {\star d{\star d\phi}} = \left(-1\right)^{t + D-1}\nabla^{\mu}\nabla_{\mu}\phi\,,
\end{equation}
\noindent while a one-form $A$ with field strength $F=dA$ satisfies
\begin{equation}
{\star d{\star dA}} = {\star d{\star F}}=\left(-1\right)^{t+1}\nabla_{\nu}F^{\nu}_{\phantom{\nu}\lambda}\,dx^{\lambda}\,.
\end{equation}

\subsection{Details of the Kaluza-Klein reduction}\label{subsec:KK}
Here we provide further details on the reduction of the $5d$ theory \eqref{eqn:Lagrangian5d b} on the two-sphere. As described in the main text, our KK ansatz is
\begin{align}
ds_{5}^{2} ={}&
 ds^{2}_{\text{string}}(M) + e^{2U(x)}ds^{2}\left(Y\right)
 \\
 \tilde{F}^{i} ={}&
-B_{i}\, \sin\theta\,d\theta\wedge d\phi
 \\
\Psi ={}&
 \Psi(x)
 \\
 \Phi = {}&
 \Phi(x)\, ,
\end{align}
\noindent where $M$ is the $(2+1)$-dimensional external manifold with coordinates $x = \{r,t,z\}$ and $Y$ is the two-sphere with radius $\ell_{S}$ and coordinates $y = \{\theta,\phi\}$. We pick the orientation such that the volume form on $Y$ is $\text{vol}_{2} =\ell_{S}^{2} \sin\theta\, d\theta\wedge d\phi\,$.

Because the field strengths $\tilde{F}^{i}$ are purely magnetic, and proportional to the volume form of the two-sphere, the vector equations \eqref{5d eom} are satisfied trivially in our ansatz and do not yield lower-dimensional equations of motion. Let us now consider the reduction of the scalar equations \eqref{5d eom 2}-\eqref{5d eom 3}. In order to reduce the coupling of the scalars to the U(1) field strengths it is useful to notice that via \eqref{square of forms} our ansatz implies
\begin{equation}\label{magnetic F squared}
{\,\star_5 \tilde{F}^i} \wedge \tilde F^i = \frac{1}{2!}\tilde{F}^{i\, \mu\nu}\tilde{F}^{i}_{\mu\nu}\,\text{vol}_{5} = \frac{B_{i}^{2}}{\ell_{S}^{4}}e^{-4U(x)}\,\text{vol}_{5} =\frac{B_{i}^{2}}{\ell_{S}^{4}} e^{-2U(x)}\, \text{vol}_{3}\wedge  \text{vol}_{2}\,.
\end{equation}
\noindent Next, we note that for any one-form $A$ with support in $M$
\begin{equation}
{\star_{5}A } =  e^{2U(x)}{\star_{3}A}\wedge \text{vol}_{2}\,.
\end{equation}
\noindent In particular, if $\Psi$ is a scalar in $M$, applying this result to $d\Psi$ we find
\begin{equation}
{\star_{5}d\Psi } =  e^{2U(x)}{\star_{3}d\Psi}\wedge \text{vol}_{2}\,.
\end{equation}
\noindent The decomposition of the scalar Laplacian then follows:
\begin{equation}
d{\star_{5}d\Psi } =e^{2U(x)}\Bigl[d\left({\star_{3}d\Psi}\right)+2dU(x)\wedge{\star_{3}d\Psi}\Bigr]\wedge \text{vol}_{2}\,.
\end{equation}
\noindent Plugging this result together with \eqref{magnetic F squared} into \eqref{5d eom 2}-\eqref{5d eom 3} we find the effective $3d$ equations for the scalar fields on $M$:
\begin{align}\label{3d Psi eq string frame}
0 ={}&
 d\left({\star_{3}d\Psi}\right)+2dU\wedge{\star_{3}d\Psi}  - \frac{e^{-4U}}{2}\sum_{i=1}^{3}\frac{B_{i}^{2}}{\ell_{S}^{4}}\frac{\delta H_{ii}}{\delta \Psi} \text{vol}_{3}
\\
0 ={}&
 d\left({\star_{3}d\Phi}\right)+2dU\wedge{\star_{3}d\Phi}  - \frac{e^{-4U}}{2}\sum_{i=1}^{3} \frac{B_{i}^{2}}{\ell_{S}^{4}}\frac{\delta H_{ii}}{\delta \Phi}\text{vol}_{3}\,.
 \label{3d Phi eq string frame}
\end{align}
\noindent Equivalently, in component notation we have
\begin{align}
0 ={}&
 \nabla_{\mu}\nabla^{\mu}\Psi +2\nabla_{\mu}U\nabla^{\mu}\Psi  - \frac{e^{-4U}}{2}\sum_{i=1}^{3} \frac{B_{i}^{2}}{\ell_{S}^{4}}\frac{\delta H_{ii}(\Psi,\Phi)}{\delta \Psi}
\\
0 ={}&
 \nabla_{\mu}\nabla^{\mu}\Phi +2\nabla_{\mu}U\nabla^{\mu}\Phi  - \frac{e^{-4U}}{2}\sum_{i=1}^{3}\frac{ B_{i}^{2}}{\ell_{S}^{4}}\frac{\delta H_{ii}(\Psi,\Phi)}{\delta \Phi} \,.
\end{align}

We now turn our attention to the reduction of the $5d$ Einstein's equations \eqref{5d Eins}. In order to reduce the Ricci tensor, we first study the decomposition of the spin connection and the curvature two-form. Let $\hat{e}^{M}$ denote the $5d$ local Lorentz frame, and $M,N,\ldots$ denote the flat indices on the $5d$ manifold. Denoting by $a,b,\ldots$ the flat indices on $M$, and by $\alpha,\beta,\ldots$ the flat indices on the compact manifold $Y$, our choice of vielbein reads
\begin{align}
\hat e^a &= e^a\\
\hat e^\alpha &= e^U e^\alpha\,,
\end{align}
\noindent where $e^{a}$ and $e^{\alpha}$ are orthonormal frames for $M$ and $Y$, respectively.  Denoting by ${\omega^a}_b$ the spin connection associated with $M$ and by ${\omega^\alpha}_\beta$ the spin connection appropriate to $Y$, solving the torsionless condition for the $5d$ spin connection $\hat{\omega}$ we find
\begin{align}
\hat{\omega}^{a}_{\hphantom{a}b} &= \omega^{a}_{\hphantom{a}b}
\\
\hat{\omega}^{\alpha}_{\hphantom{\alpha}\beta} &= \omega^{\alpha}_{\hphantom{\alpha}\beta}
\\
\hat{\omega}^{\alpha}_{\hphantom{\alpha}a} &= P_{a}\,e^\alpha\,,
\end{align}
\noindent where we introduced the shorthand
\begin{equation}
P_{a} \equiv   e^{U}(\partial_a U)\,.
\end{equation}
\noindent It is useful to notice that  $\hat{\omega}^{a}_{\hphantom{a}\alpha} \wedge \hat{\omega}^{\alpha}_{\hphantom{\alpha}b} = P^{a} P_{b}\,\eta_{\alpha\beta}\,e^{\alpha}\wedge e^{\beta}= 0$. Next, let $\Theta$ denote the curvature two-form. Then, on the $5d$ manifold we have $\hat{\Theta}^{M}_{\hphantom{M}N} = d\hat{\omega}^{M}_{\hphantom{M}N} + \hat{\omega}^{M}_{\hphantom{M}P}\wedge \hat{\omega}^{P}_{\hphantom{P}N}\,$. Computing the different components we find
\begin{align}
\hat{\Theta}^{a}_{\hphantom{a}b} ={}&
\Theta^{a}_{\hphantom{a}b}
\\
\hat{\Theta}^{\alpha}_{\hphantom{\alpha}\beta} ={}&
\Theta^{\alpha}_{\hphantom{\alpha}\beta} - P_{a}P^{a}\eta_{\beta [\gamma}\delta^{\alpha}_{\hphantom{\alpha}\sigma]}\, e^{\sigma}\wedge e^{\gamma}
\\
\hat{\Theta}^{\alpha}_{\hphantom{\alpha}a}
={}&
\delta^{\alpha}_{\hphantom{\alpha}\gamma}\left(\nabla_{c}P_{a}\right) e^{c}\wedge e^{\gamma}\,.
\end{align}
\noindent The antisymmetrization symbol $[...]$ used above includes a factor of $1/2!\,$, and $\nabla_{a}$ denotes the connection on $M$. From these expressions we can identify the non-vanishing components of the Riemann tensor, defined as $\hat{\Theta}^{M}_{\hphantom{M}N} = \frac{1}{2!}\hat{R}^{M}_{\hphantom{M}NPQ}\,\hat{e}^{P}\wedge\hat{e}^{Q}\,$:
\begin{align}
\hat{R}^{a}_{\hphantom{a}bcd}
={}&
 R^{a}_{\hphantom{a}bcd}
 \\
 \hat{R}^{\alpha}_{\hphantom{\alpha}\beta\gamma\delta}
={}&
 e^{-2U}\mathcal{R}^{\alpha}_{\hphantom{\alpha}\beta \gamma \delta} -2e^{-2U}P_{a}P^{a}\delta^{\alpha}_{\hphantom{\alpha}[\gamma}\eta_{\delta]\beta}
 \\
  \hat{R}^{\alpha}_{\hphantom{\alpha}a \beta b}
  ={}&
   -\delta^{\alpha}_{\hphantom{\alpha}\beta}\,e^{-U}\nabla_{b}P_{a}
   \\
     \hat{R}^{a}_{\hphantom{a}\alpha b \beta }
     ={}&
     -\eta_{\alpha\beta}\,e^{-U}\nabla_{b}P^{a}\,.
\end{align}
\noindent In the above notation $R^{a}_{\hphantom{a}bcd}$ are the components of the Riemann tensor of the external manifold $M$, and $\mathcal{R}^{\alpha}_{\hphantom{\alpha}\beta \gamma \delta}$ those of the Riemann tensor of the compact manifold $Y$. Finally, for the decomposition of the Ricci tensor $\hat{R}_{MN} = \hat{R}^{P}_{\hphantom{P}MPN}$ we find
\begin{align}
\hat{R}_{ab} ={}&
 R_{ab} - d_{Y}e^{-U}\nabla_{b}P_{a}
\nonumber\\
={}&
R_{ab} - d_{Y}\left(\nabla_{b}\nabla_{a}U + \nabla_{a}U\nabla_{b}U\right)
\\
\hat{R}_{\alpha\beta} ={}&
e^{-2U}\mathcal{R}_{\alpha\beta} - \left(d_{Y}-1\right)e^{-2U}P_{a}P^{a}\,\eta_{\alpha\beta} -e^{-U}\nabla_{c}P^{c}\,\eta_{\alpha\beta}
\nonumber\\
={}&
e^{-2U}\mathcal{R}_{\alpha\beta} - d_{Y}\left(\nabla_{a}U\nabla^{a}U\right)\eta_{\alpha\beta} - \left(\nabla^{a}\nabla_{a}U\right)\eta_{\alpha\beta}
\\
\hat{R}_{a\alpha} ={}& 0\,,
\end{align}
\noindent where $d_{Y}$ is the dimension of the compact manifold ($d_{Y}=2$ in our case). In particular, for the Ricci scalar $\hat{R} = \eta^{MN}\hat{R}_{MN}$ it follows that
\begin{align}
\hat{R}
 ={}&
  R+ e^{-2U}\mathcal{R} - 2d_{Y}\nabla^{a}\nabla_{a}U  - d_{Y}\left(1+ d_{Y}\right) \nabla^{a}U\nabla_{a}U\,,
\end{align}
\noindent where $R$ is the scalar curvature on $M$, and $\mathcal{R}$ that of $Y$. Since the two-sphere has radius $\ell_{S}$ we have $\mathcal{R}_{\alpha\beta}=\eta_{\alpha\beta}/ \ell_{S}^{2}\,$. Setting $d_{Y}=2$ in the above expressions we find that the only non-vanishing components in our reduction are
\begin{align}
\hat{R}_{ab} ={}&
R_{ab} - 2\bigl(\nabla_{b}\nabla_{a}U + \nabla_{a}U\nabla_{b}U\bigr)
\\
\hat{R}_{\alpha\beta} ={}&
\biggl(\frac{e^{-2U}}{\ell_{S}^{2}}  - \nabla^{a}\nabla_{a}U- 2\nabla_{a}U\nabla^{a}U\biggr)\eta_{\alpha\beta}\,.
\end{align}
\noindent The Ricci scalar is then given by
\begin{equation}
\hat{R} = R+ \frac{2}{\ell_{S}^{2}}e^{-2U} - 4\nabla^{a}\nabla_{a}U  - 6 \nabla^{a}U\nabla_{a}U\,.
\end{equation}
\noindent Using the decomposition of the Ricci tensor, from the components of the $5d$ Einstein's equations in the directions of the external manifold $M$ we get (using flat indices on $M$)
\begin{align}\label{3d Einstein eq string frame}
R_{ab}
 ={}&
 2\bigl(\nabla_{b}\nabla_{a}U + \nabla_{a}U\nabla_{b}U\bigr)
 \nonumber\\
 &
 +\frac{1}{2}\biggl[\nabla_{a}\Psi\nabla_{b}\Psi +\nabla_{a}\Phi\nabla_{b}\Phi
 -  \frac{\eta_{ab}}{3}e^{-4U}\sum_{i=1}^{3} \frac{B_{i}^{2}}{\ell_{S}^{4}}H_{ii}(\Psi,\Phi)\biggr].
\end{align}
\noindent Similarly, noting that with flat indices $\tilde{F}^{i\hphantom{i}P}_{\alpha}\tilde{F}^{i}_{\beta P} = \left(e^{-4U}B_{i}^{2}/\ell_{S}^{4}\right)\eta_{\alpha\beta}\,$, from the components of the $5d$ Einstein's equations in the directions of $Y$ we find
\begin{equation}\label{U equation string frame}
      \nabla^{a}\nabla_{a}U+ 2\nabla_{a}U\nabla^{a}U-\frac{e^{-2U}}{\ell_{S}^{2}} +\frac{e^{-4U}}{3}\sum_{i=1}^{3}\frac{B_{i}^{2}}{\ell_{S}^{4}}H_{ii}(\Psi,\Phi) = 0\,.
\end{equation}
\noindent Since all reference to the two-sphere dropped out from the equations of motion, the proposed truncation is consistent. Finally, we point out that the resulting three-dimensional equations of motion \eqref{3d Psi eq string frame}, \eqref{3d Phi eq string frame}, \eqref{3d Einstein eq string frame} and \eqref{U equation string frame} can be obtained from the following effective action (in string frame):
\begin{align}
S_{\text{string}} = -\frac{1}{16\pi G_3}\int d^{3}x \sqrt{|g|}\,e^{2U}\biggl[&
R + \frac{2}{\ell_{S}^{2}}e^{-2U} - \frac{e^{-4U}}{2}\sum_{i=1}^{3}\frac{B_{i}^{2}}{\ell_{S}^{4}}H_{ii}(\Psi,\Phi)
\nonumber\\
&
+ 2\left(\nabla U\right)^{2} -\frac{1}{2}\left(\nabla\Psi\right)^{2}-\frac{1}{2}\left(\nabla\Phi\right)^{2} \biggr].
\end{align}

\subsection{The general solution in terms of $5d$ and $3d$ fields} \label{subsec:scalars}
 In terms of the $4d$ fields \eqref{4d ansatz 1}-\eqref{4d ansatz 4}, our four-parameter family of solutions was given in \eqref{general solution 1}-\eqref{general solution 4} (with $F^{0}$ given by \eqref{field strengths on shell}). The $h^{i}$ fields appearing in the $5d$ theory (and also in the $4d$ STU model) are related to the diagonal fields \eqref{diagonal fields 1}-\eqref{diagonal fields 2} through
 \begin{align}
h^{1} ={}&
 \exp\left[\frac{1}{6}\left(2\phi_{1}-\phi_{2}-\phi_{3}\right)\right]
\\
h^{2} ={}&
 \exp\left[\frac{1}{6}\left(2\phi_{2}-\phi_{1}-\phi_{3}\right)\right]
\\
h^{3} ={}&
 \exp\left[\frac{1}{6}\left(2\phi_{3}-\phi_{1}-\phi_{2}\right)\right].
\end{align}
\noindent Hence, in our general family of solutions \eqref{gen regular solution 1}-\eqref{gen regular solution 2} they read
\begin{align}
\label{general h1}
h^{1}(r) ={}&
 \left(\frac{B_{1}^{2}}{|B_{2}B_{3}|}\frac{\sqrt{\left(1+a_{2}^{2}\right)\left(1+a_{3}^{2}\right)}}{1+a_{1}^{2}}\frac{\left(a_{1}^{2}\,r + 2m\right)^{2}}{\left(a_{2}^{2}\,r+2m\right)\left(a_{3}^{2}\,r + 2m\right)}\right)^{1/3}
 \\
h^{2}(r) ={}&
 \left(\frac{B_{2}^{2}}{|B_{1}B_{3}|}\frac{\sqrt{\left(1+a_{1}^{2}\right)\left(1+a_{3}^{2}\right)}}{1+a_{2}^{2}}\frac{\left(a_{2}^{2}\,r + 2m\right)^{2}}{\left(a_{1}^{2}\,r+2m\right)\left(a_{3}^{2}\,r + 2m\right)}\right)^{1/3}
  \\
h^{3}(r) ={}&
 \left(\frac{B_{3}^{2}}{|B_{1}B_{2}|}\frac{\sqrt{\left(1+a_{1}^{2}\right)\left(1+a_{2}^{2}\right)}}{1+a_{3}^{2}}\frac{\left(a_{3}^{2}\,r + 2m\right)^{2}}{\left(a_{1}^{2}\,r+2m\right)\left(a_{2}^{2}\,r + 2m\right)}\right)^{1/3} \,.
 \label{general h3}
\end{align}
\noindent We recall that the $5d$ line element is given in terms of the $4d$ one by \eqref{uplifted line element}. Similarly, in terms of the decoupled fields the $3d$ scalars are given by
\begin{align}
U ={}&
  \frac{\phi_{1}+\phi_{2}+\phi_{3}}{6} + \log \frac{m}{\ell_{S}}
 \\
\Psi ={}&
\frac{1}{2\sqrt{6}}\left(2\phi_{1} -\phi_{2}-\phi_{3}\right)
\\
\Phi ={}&
\frac{1}{2\sqrt{2}}\left(\phi_{3}-\phi_{2}\right)\,,
\end{align}
\noindent so in our general solution we obtain
\begin{align}\label{general U}
U(r) ={}&
\frac{1}{3}\log\Biggl[\frac{1}{\sqrt{\left(1 + a_{1}^{2}\right)\left(1 + a_{2}^{2}\right)\left(1 + a_{3}^{2}\right)}}\left(\frac{a_{1}^{2}}{2m}r + 1\right)\left(\frac{a_{2}^{2}}{2m}r + 1\right)\left(\frac{a_{3}^{2}}{2m}r + 1\right)\Biggr]
\\
\Psi(r) ={}&
\frac{1}{\sqrt{6}}\log\left[\frac{B_{1}^{2}}{|B_{2}B_{3}|}\frac{\sqrt{\left(1+a_{2}^{2}\right)\left(1+a_{3}^{2}\right)}}{1+a_{1}^{2}}\frac{\left(a_{1}^{2}\,r + 2m\right)^{2}}{\left(a_{2}^{2}\, r+2m\right)\left(a_{3}^{2}\,r+2m\right)}\right]
\\
\Phi(r) ={}&
\frac{1}{\sqrt{2}}\log\left[\left|\frac{B_{3}}{B_{2}}\right|\sqrt{\frac{1+a_{2}^{2}}{1+a_{3}^{2}}}\left(\frac{a_{3}^{2}\,r + 2m}{a_{2}^{2}\,r +2m}\right)\right].
\label{general Phi}
\end{align}

\newpage

\begin{thebibliography}{10}

\bibitem{Strominger:1996sh}
A.~Strominger and C.~Vafa, {\it {Microscopic origin of the Bekenstein-Hawking
  entropy}},  {\em Phys.Lett.} {\bf B379} (1996) 99--104,
  [\href{http://xxx.lanl.gov/abs/hep-th/9601029}{{\tt hep-th/9601029}}].

\bibitem{David:2002wn}
J.~R. David, G.~Mandal, and S.~R. Wadia, {\it {Microscopic formulation of black
  holes in string theory}},  {\em Phys.Rept.} {\bf 369} (2002) 549--686,
  [\href{http://xxx.lanl.gov/abs/hep-th/0203048}{{\tt hep-th/0203048}}].

\bibitem{Kraus:2006wn}
P.~Kraus, {\it {Lectures on black holes and the AdS(3) / CFT(2)
  correspondence}},  {\em Lect.Notes Phys.} {\bf 755} (2008) 193--247,
  [\href{http://xxx.lanl.gov/abs/hep-th/0609074}{{\tt hep-th/0609074}}].

\bibitem{Sen:2007qy}
A.~Sen, {\it {Black Hole Entropy Function, Attractors and Precision Counting of
  Microstates}},  {\em Gen.Rel.Grav.} {\bf 40} (2008) 2249--2431,
  [\href{http://xxx.lanl.gov/abs/0708.1270}{{\tt arXiv:0708.1270}}].

\bibitem{Guica:2008mu}
M.~Guica, T.~Hartman, W.~Song, and A.~Strominger, {\it {The Kerr/CFT
  Correspondence}},  {\em Phys.Rev.} {\bf D80} (2009) 124008,
  [\href{http://xxx.lanl.gov/abs/0809.4266}{{\tt arXiv:0809.4266}}].

\bibitem{Hartman:2008pb}
T.~Hartman, K.~Murata, T.~Nishioka, and A.~Strominger, {\it {CFT Duals for
  Extreme Black Holes}},  {\em JHEP} {\bf 0904} (2009) 019,
  [\href{http://xxx.lanl.gov/abs/0811.4393}{{\tt arXiv:0811.4393}}]. 18 pages.

\bibitem{Compere:2012jk}
G.~Compere, {\it {The Kerr/CFT correspondence and its extensions: a
  comprehensive review}},  \href{http://xxx.lanl.gov/abs/1203.3561}{{\tt
  arXiv:1203.3561}}.

\bibitem{Cvetic:1997uw}
M.~Cvetic and F.~Larsen, {\it {General rotating black holes in string theory:
  Grey body factors and event horizons}},  {\em Phys.Rev.} {\bf D56} (1997)
  4994--5007, [\href{http://xxx.lanl.gov/abs/hep-th/9705192}{{\tt
  hep-th/9705192}}].

\bibitem{Maldacena:1996ix}
J.~M. Maldacena and A.~Strominger, {\it {Black hole grey body factors and
  d-brane spectroscopy}},  {\em Phys.Rev.} {\bf D55} (1997) 861--870,
  [\href{http://xxx.lanl.gov/abs/hep-th/9609026}{{\tt hep-th/9609026}}].

\bibitem{Cvetic:2009jn}
M.~Cvetic and F.~Larsen, {\it {Greybody Factors and Charges in Kerr/CFT}},
  {\em JHEP} {\bf 0909} (2009) 088,
  [\href{http://xxx.lanl.gov/abs/0908.1136}{{\tt arXiv:0908.1136}}].

\bibitem{Castro:2010fd}
A.~Castro, A.~Maloney, and A.~Strominger, {\it {Hidden Conformal Symmetry of
  the Kerr Black Hole}},  {\em Phys.Rev.} {\bf D82} (2010) 024008,
  [\href{http://xxx.lanl.gov/abs/1004.0996}{{\tt arXiv:1004.0996}}].

\bibitem{Cvetic:2011hp}
M.~Cvetic and F.~Larsen, {\it {Conformal Symmetry for General Black Holes}},
  {\em JHEP} {\bf 1202} (2012) 122,
  [\href{http://xxx.lanl.gov/abs/1106.3341}{{\tt arXiv:1106.3341}}].

\bibitem{Cvetic:2011dn}
M.~Cvetic and F.~Larsen, {\it {Conformal Symmetry for Black Holes in Four
  Dimensions}},  \href{http://xxx.lanl.gov/abs/1112.4846}{{\tt
  arXiv:1112.4846}}.

\bibitem{Cvetic:2012tr}
M.~Cvetic and G.~Gibbons, {\it {Conformal Symmetry of a Black Hole as a Scaling
  Limit: A Black Hole in an Asymptotically Conical Box}},  {\em JHEP} {\bf
  1207} (2012) 014, [\href{http://xxx.lanl.gov/abs/1201.0601}{{\tt
  arXiv:1201.0601}}].

\bibitem{Cremmer:1984hj}
E.~Cremmer, C.~Kounnas, A.~Van~Proeyen, J.~Derendinger, S.~Ferrara, et~al.,
  {\it {Vector Multiplets Coupled to N=2 Supergravity: SuperHiggs Effect, Flat
  Potentials and Geometric Structure}},  {\em Nucl.Phys.} {\bf B250} (1985)
  385.

\bibitem{Duff:1995sm}
M.~Duff, J.~T. Liu, and J.~Rahmfeld, {\it {Four-dimensional
  string-string-string triality}},  {\em Nucl.Phys.} {\bf B459} (1996)
  125--159, [\href{http://xxx.lanl.gov/abs/hep-th/9508094}{{\tt
  hep-th/9508094}}].

\bibitem{Behrndt:1996hu}
K.~Behrndt, R.~Kallosh, J.~Rahmfeld, M.~Shmakova, and W.~K. Wong, {\it {STU
  black holes and string triality}},  {\em Phys.Rev.} {\bf D54} (1996)
  6293--6301, [\href{http://xxx.lanl.gov/abs/hep-th/9608059}{{\tt
  hep-th/9608059}}].

\bibitem{Virmani:2012kw}
A.~Virmani, {\it {Subtracted Geometry From Harrison Transformations}},  {\em
  JHEP} {\bf 1207} (2012) 086, [\href{http://xxx.lanl.gov/abs/1203.5088}{{\tt
  arXiv:1203.5088}}].

\bibitem{Huijse:2011ef}
L.~Huijse, S.~Sachdev, and B.~Swingle, {\it {Hidden Fermi surfaces in
  compressible states of gauge-gravity duality}},  {\em Phys.Rev.} {\bf B85}
  (2012) 035121, [\href{http://xxx.lanl.gov/abs/1112.0573}{{\tt
  arXiv:1112.0573}}].

\bibitem{Dong:2012se}
X.~Dong, S.~Harrison, S.~Kachru, G.~Torroba, and H.~Wang, {\it {Aspects of
  holography for theories with hyperscaling violation}},  {\em JHEP} {\bf 1206}
  (2012) 041, [\href{http://xxx.lanl.gov/abs/1201.1905}{{\tt
  arXiv:1201.1905}}].

\bibitem{Hartnoll:2012wm}
S.~A. Hartnoll and E.~Shaghoulian, {\it {Spectral weight in holographic scaling
  geometries}},  {\em JHEP} {\bf 1207} (2012) 078,
  [\href{http://xxx.lanl.gov/abs/1203.4236}{{\tt arXiv:1203.4236}}].

\end{thebibliography}
\providecommand{\href}[2]{#2}\begingroup\raggedright\endgroup
\end{document}